\documentclass{article}
\usepackage{fullpage}

\usepackage{amsmath}
\usepackage{amsthm} 
\usepackage{amssymb}

\usepackage{hyperref}
\usepackage{nonfloat}
\usepackage{rotating}
\usepackage{url,xspace}
\usepackage{bussproofs}
\usepackage{mathbbol}
\usepackage[all]{xy}
\SelectTips{cm}{}

\theoremstyle{plain}
\newtheorem{thm}{Theorem}[section]
\newtheorem{prop}[thm]{Proposition}
\theoremstyle{definition}
\newtheorem{defi}[thm]{Definition}
\newtheorem{exa}[thm]{Example}
\newtheorem{rem}[thm]{Remark}
\newtheorem{claim}[thm]{Claim}

\newcommand{\iso}{\cong}

\newcommand{\lto}{\leftarrow}
\newcommand{\rupto}[1]{\stackrel{#1}{\rightarrow}}
\newcommand{\lupto}[1]{\stackrel{#1}{\leftarrow}}
\newcommand{\deno}[1]{[[#1]]}

\newcommand{\bank}{\mathrm{bank}}

\newcommand{\pure}{{(0)}} 
\newcommand{\eqs}{\equiv} 
\newcommand{\eqw}{\sim} 
\newcommand{\ti}{\widetilde}
\newcommand{\ha}{\widehat}

\newcommand{\unit}{\mathbb{1}}
\newcommand{\acc}{{(1)}} 
\newcommand{\modi}{{(2)}} 
 
\newcommand{\Loc}{\mathit{Loc}} 
\newcommand{\Val}{\mathit{Val}} 
\newcommand{\St}{\mathit{St}} 
\newcommand{\prl}{\mathit{prl}} 
\newcommand{\prr}{\mathit{prr}} 
\newcommand{\fl}{l} 
\newcommand{\fu}{u} 
\newcommand{\sfl}{l} 
\newcommand{\sfu}{u} 
\newcommand{\tuple}[1]{\langle #1 \rangle}
\newcommand{\tu}{\tuple{\,}}

\newcommand{\empt}{\mathbb{0}}
\newcommand{\ppg}{{(1)}} 
\newcommand{\ctc}{{(2)}} 
 
\newcommand{\Const}{\mathit{ExCstr}} 
\newcommand{\Par}{\mathit{Par}} 
\newcommand{\Exc}{\mathit{Exc}} 
\newcommand{\inl}{\mathit{inl}} 
\newcommand{\inr}{\mathit{inr}} 
\newcommand{\ft}{t} 
\newcommand{\fc}{c} 
\newcommand{\sft}{t} 
\newcommand{\sfc}{c} 
\newcommand{\cotuple}[1]{[ #1 ]}
\newcommand{\cotu}{\cotuple{\,}}
\newcommand{\toppg}[1]{\triangledown #1 } 
\newcommand{\Ppg}{\mathit{Ppg}} 
\newcommand{\throw}[2]{\mathit{throw}_{#1,#2}} 
\newcommand{\try}[2]{\mathit{try}\{#1\}\,#2}  
\newcommand{\catch}[2]{\mathit{catch}\,#1\,\{#2\}}  
\newcommand{\catchrec}[3]{\mathit{catch}\,#1\,\{#2\}\,#3}  
\newcommand{\catchn}[1]{\mathit{catch}\,#1}  
\newcommand{\rais}[2]{\mathit{raise}_{#1,#2}}  
\newcommand{\handle}[3]{#1\;\mathit{handle}\;#2\!\Rightarrow\!#3}  
\newcommand{\handlerec}[5]{#1\;\mathit{handle}
\;#2\!\Rightarrow\!#3\,|\,\dots\,|\,#4\!\Rightarrow\!#5}  
\newcommand{\handlen}[2]{#1\;\mathit{handle}\;#2}  
\newcommand{\Handle}{H}  

\newcommand{\copi}{\iota} 
\newcommand{\lplus}{+} 
\newcommand{\rplus}{+} 

\newcommand{\Ss}{\Sigma} 
\newcommand{\Ff}{\Phi} 
\newcommand{\Tt}{\Theta} 
\newcommand{\Hh}{\mathcal{H}} 
\newcommand{\Cc}{\mathcal{C}} 
\renewcommand{\ss}{\sigma}

\newcommand{\eqn}{\mathrm{eq}} 
\newcommand{\st}{\mathrm{st}} 
\newcommand{\exc}{\mathrm{exc}} 
 
\newcommand{\inte}{\mathrm{int}} 
\newcommand{\set}{\mathrm{set}} 
\newcommand{\deco}{\mathrm{deco}} 
\newcommand{\app}{\mathrm{app}} 
\newcommand{\expl}{\mathrm{expl}} 
\newcommand{\all}{\mathrm{all}} 
\newcommand{\ob}{\mathrm{obs}} 
\newcommand{\encaps}{\mathrm{encaps}} 

\newcommand{\bZ}{\mathbb{Z}} 
\newcommand{\bU}{\{\star\}} 
\newcommand{\bN}{\mathbb{N}} 

\newcommand{\catC}{\mathbf{C}}
\newcommand{\catS}{\mathbf{S}} 
\newcommand{\catT}{\mathbf{T}} 

\newcommand{\rn}[1]{\;\textrm{(#1)}\;} 
\newcommand{\rncomp}{\rn{comp}}
\newcommand{\rnid}{\rn{id}}
\newcommand{\rnassoc}{\rn{assoc}}
\newcommand{\rnidsrc}{\rn{id-src}}
\newcommand{\rnidtgt}{\rn{id-tgt}}
\newcommand{\rnsrefl}{\rn{$\eqs$-refl}}
\newcommand{\rnssym}{\rn{$\eqs$-sym}}
\newcommand{\rnstrans}{\rn{$\eqs$-trans}}
\newcommand{\rnssubs}{\rn{$\eqs$-subs}}
\newcommand{\rnsrepl}{\rn{$\eqs$-repl}}
\newcommand{\rnpa}{\rn{0-to-1}}
\newcommand{\rnam}{\rn{1-to-2}}
\newcommand{\rnpcomp}{\rn{0-comp}}
\newcommand{\rnacomp}{\rn{1-comp}}
\newcommand{\rnpid}{\rn{0-id}}
\newcommand{\rnwrefl}{\rn{$\eqw$-refl}}
\newcommand{\rnwsym}{\rn{$\eqw$-sym}}
\newcommand{\rnwtrans}{\rn{$\eqw$-trans}}
\newcommand{\rnwsubs}{\rn{$\eqw$-subs}}
\newcommand{\rnwrepl}{\rn{$\eqw$-repl}}
\newcommand{\rnsw}{\rn{$\eqs$-to-$\eqw$}}
\newcommand{\rnws}{\rn{$\eqw$-to-$\eqs$}}
\newcommand{\rnfinal}{\rn{final}}
\newcommand{\rnpfinal}{\rn{0-final}}  
\newcommand{\rnwfinal}{\rn{$\eqw$-final}}
\newcommand{\rnsfinal}{\rn{$\eqs$-final}}
\newcommand{\rntuple}{\rn{tuple}}
\newcommand{\rntupleuns}{\rn{$\eqs$-tuple}}

\newcommand{\an}[1]{\;\textrm{#1}\;}
\newcommand{\anliui}{\an{A1}}
\newcommand{\anliuj}{\an{A2}}
\newcommand{\anprodpure}{\an{P1}}
\newcommand{\anprodmodi}{\an{P2}}

\newcommand{\pn}[1]{\;\textrm{#1}\;}
\newcommand{\pna}{\pn{Pr$_1$}}
\newcommand{\pnb}{\pn{Pr$_2$}}
\newcommand{\pnc}{\pn{Pr$_3$}}
\newcommand{\pnd}{\pn{Pr$_4$}}
\newcommand{\pne}{\pn{Pr$_5$}}
\newcommand{\pnf}{\pn{Pr$_6$}}
\newcommand{\png}{\pn{Pr$_7$}}
\newcommand{\pnh}{\pn{Pr$_8$}}

\newcommand{\hsp}{\null\hspace{5pt}}

\newcommand{\cpp}{\texttt{C++}\xspace}
\newcommand{\java}{\texttt{Java}\xspace}

\newcommand{\Mod}{\mathrm{Mod}}
\newcommand{\id}{\mathit{id}}
\newcommand{\zoom}{\mathcal{Z}} 
\newcommand{\Lang}{\Lambda} 

\title{States and exceptions \\ considered as dual effects}
\date{May 19., 2011}
\author
{
  Jean-Guillaume Dumas\thanks{
    LJK, Universit\'e de Grenoble, France. \url{Jean-Guillaume.Dumas@imag.fr}},
  Dominique Duval\thanks{
    LJK, Universit\'e de Grenoble, France. \url{Dominique.Duval@imag.fr}},
  Laurent Fousse\thanks{
    LJK, Universit\'e de Grenoble, France. \url{Laurent.Fousse@imag.fr}}, 
  Jean-Claude Reynaud\thanks{
    Malhivert, Claix, France. \url{Jean-Claude.Reynaud@imag.fr}}
}	

\begin{document}

\maketitle

\begin{abstract}
\textbf{Abstract.} 
In this paper we consider the two major computational effects 
of states and exceptions,
from the point of view of diagrammatic logics. 
We get a surprising result: there exists 
a symmetry between these two effects,
based on the well-known categorical duality between 
products and coproducts.
More precisely, 
the lookup and update operations for states are respectively 
dual to the throw and catch operations for exceptions. 
This symmetry is deeply hidden in the programming languages;  
in order to unveil it, we start from the monoidal equational logic 
and we add progressively the logical features which are necessary 
for dealing with either effect. 
This approach gives rise to a new point of view on states
and exceptions, which bypasses the problems due to 
the non-algebraicity of handling exceptions.  
\end{abstract}

\section*{Introduction}

In this paper we consider two major computational effects: 
states and exceptions. 
We get a surprising result: there exists 
a symmetry between these two effects,
based on the well-known categorical duality between 
products and coproducts (or sums). 

In order to get these results we use the categorical approach 
of diagrammatic logics, as introduced in \cite{Du03}
and developed in \cite{DD10}. 
For instance, in \cite{DDR11} this approach is used 
for studying an issue related to computational effects: 
controling the order of evaluation of the arguments of a function.
This paper provides one more application
of diagrammatic logics to computational effects;
a preliminary approach can be found in \cite{DR05}. 

To our knowledge, the first categorical treatment of 
computational effects is due to Moggi \cite{Mo89,Mo91};
this approach relies on monads, it is implemented 
in the programming language Haskell \cite{Wa92,Haskell}. 
Although monads are not used in this paper, 
the basic ideas underlying our approach rely on Moggi's remarks 
about notions of computations and monads. 
In view of comparing Moggi's approach and ours, 
let us quote \cite[section~1]{Mo91}.
\textit{
The basic idea behind the categorical semantics below is that, 
in order to interpret a programming
language in a category $\catC$, we distinguish the object $A$ of values 
(of type $A$) 
from the object $T A$ of computations (of type $A$), and take as denotations 
of programs (of type $A$) the elements of $T A$.
In particular, we identify the type $A$ with the object of values 
(of type $A$) 
and obtain the object of computations (of type $A$) by applying 
an unary type-constructor $T$ to $A$. We call $T$ a notion
of computation, since it abstracts away from the type of values 
computations may produce. There
are many choices for $T A$ corresponding to different notions of computations.
[\dots]
Since the denotation of programs of type $B$
are supposed to be elements of $T B$, programs of type $B$ 
with a parameter of type $A$ ought to be
interpreted by morphisms with codomain $T B$, 
but for their domain there are two alternatives, either
$A$ or $T A$, depending on whether parameters of type $A$ 
are identified with values or computations of type $A$. 
We choose the first alternative, because it entails the second. 
Indeed computations of type $A$ are the same as values of type $T A$. 
}
The examples proposed by Moggi include 
the side-effects monad $T A = (A\times S)^S$ where $S$ is the set of states 
and the exceptions monad $T A = A + E$ where $E$ is the set of exceptions. 

Later on, using the correspondence between monads
and algebraic theories, Plotkin and Power proposed to 
use Lawvere theories for dealing with 
the operations and equations related to computational effects  
\cite{PP02,HP07}. 
The operations \emph{lookup} and \emph{update} 
are related to states, 
and the operations \emph{raise} and \emph{handle} are 
related to exceptions. 
In this framework, an operation is called \emph{algebraic} 
when it satisfies some relevant genericity properties.
It happens that \emph{lookup}, \emph{update} and \emph{raise} 
are algebraic, while \emph{handle} is not \cite{PP03}. 
It follows that the handling of exceptions is quite difficult 
to formalize in this framework;
several solutions are proposed in \cite{SM04,Le06,PP09}.
In these papers, the duality between states and exceptions 
does not show up.  
One reason might be that, as we will see in this paper, 
exceptions catching is encapsulated in several nested conditionals 
which hide this duality.  

Let us look more closely at the monad of exceptions $T A = A + E$.
According to the point of view of monads for effects, 
a morphism from $A$ to $TB$ provides a 
denotation for a program of type $B$
with a parameter of type $A$.
Such a program may raise an exception,
by mapping some $a\in A$ to an exception $e\in E$.  
In order to catch an exception, it should also be possible to 
map some $e\in E$ to a non-exceptional value $b\in B$.  
We formalize this property by choosing 
the second alternative in Moggi's discussion: 
programs of type $B$ with a parameter of type $A$ are 
interpreted by morphisms with codomain $T B$ 
\textit{and with domain $T A$},
where the elements of $T A$ are seen as  
computations of type $A$ rather than values of type $T A$.
This example enlightens one of the reasons why we generalize Moggi's approach. 
What is kept, and even emphasized,
is the distinction between several kinds of programs. 
In fact, for states as well as for exceptions, we  
distinguish three kinds of programs, and moreover two kinds of equations. 
A computational effect is seen as an \emph{apparent lack of soundness}: 
the intended denotational semantics is not sound,
in the sense that it does not satisfy the given axioms, 
however it becomes sound when some additional information is given.

In order to focus on the effects,
our study of states and exceptions is based on 
a very simple logic: the monadic equational logic. 
First we provide a detailed description of the intended 
denotational \emph{semantics} of states and exceptions, 
using explicitly a set of states and a set of exceptions
(claims~\ref{claim:states-explicit} and~\ref{claim:exceptions-explicit}).
The duality between states and exceptions derives 
in an obvious way from our presentation
(proposition~\ref{proposition:duality-explicit}).
It is a duality between the lookup and update
operations for states, on one hand,
and the \emph{key} throwing and catching 
operations for exceptions, on the other hand. 
The key part in throwing an exception
is the mapping of some non-exceptional value to an exception,
while the key part in catching an exception
is the mapping of some exception to a non-exceptional value.  
Then these key operations have to be encapsulated 
in order to get the usual raising and handling of exceptions: 
handling exceptions is obtained by
encapsulating the key catching operation inside conditionals. 
Then we describe the \emph{syntax} of states and exceptions. 
The computational effects lie in the fact that 
this syntax does not mention any ``type of states'' or ``type of exceptions'', 
respectively. 
There are two variants for this syntax: 
the intended semantics is not a model of the \emph{apparent syntax}, 
but this lack of soundness is fixed in the \emph{decorated syntax} 
by providing some additional information 
(propositions~\ref{proposition:states} and~\ref{proposition:exceptions}).
The duality between states and the key part of exceptions 
holds at the syntax level as a duality of effects
(theorem~\ref{theorem:duality}),
from which the duality at the semantics level derives easily.
We use three different logics for formalizing each computational effect:
the intended semantics is described in the \emph{explicit} logic, 
the apparent syntax in the \emph{apparent} logic
and the decorated syntax in the \emph{decorated} logic. 
The explicit and apparent logics are ``usual'' logics; 
in order to focus on the effects we choose two variants of
the monadic equational logic. 
The framework of \emph{diagrammatic logics} provides a simple  
description of the three logics,
including the ``unusual'' decorated logic;
most importantly, it provides  
a relevant notion of morphisms for relating these three logics. 

The paper is organized as follows.
The intended semantics of states and exceptions is given 
in section~\ref{sec:explicit},
and the duality is described at the semantics level.
Then a simplified version of the framework of diagrammatic logics 
for effects is presented 
in section~\ref{sec:effects}, together with a motivating example 
in order to introduce the notion of ``decoration''. 
Section~\ref{sec:states} is devoted to states 
and section~\ref{sec:exceptions} to exceptions.
In section~\ref{sec:dual}, the duality is extended to the syntax level. 
In appendix~\ref{app}, some fundamental properties of states and exceptions
are proved in the decorated logic. 
In this paper, the word ``apparent'' is used in the sense of ``seeming''
(``appearing as such but not necessarily so'').  

\section{States and exceptions: duality of denotational semantics} 
\label{sec:explicit}

In this section, the symmetry between states and exceptions 
is presented 
as a duality between their intended denotational semantics
(proposition~\ref{proposition:duality-explicit}).
The aim of the next sections is to extend this result so as to get  
a symmetry between the syntax of states and exceptions, 
considered as computational effects, from which the 
duality between their semantics can be derived
(theorem~\ref{theorem:duality}). 
In this section we are dealing with sets and functions; 
the symbols $\times$ and $\prod$ are used for cartesian products, 
$+$ and $\sum$ for disjoint unions;
cartesian products are \emph{products} in the category of sets
and disjoint unions are \emph{sums} or \emph{coproducts} in this category.

\subsection{States}
\label{subsec:explicit-states}

Let $\St$ denote the set of \emph{states}. 
Let $\Loc$ denote the set of \emph{locations}
(also called \emph{variables} or \emph{identifiers}). 
For each location $i$, 
let $\Val_i$ denote the set of possible \emph{values} for $i$.
For each $i\in\Loc$ there is a \emph{lookup} function $\sfl_i:\St\to\Val_i$ 
for reading the value of location $i$ in the given state.
In addition, for each $i\in\Loc$ there is
an \emph{update} function $\sfu_i:\Val_i\times\St\to\St$ for  
setting the value of location $i$ to the given value,
without modifying the values of the other locations in the given state. 
This is summarized as follows. For each $i\in\Loc$ there are:
\begin{itemize}
\item a set $\Val_i$ (values) 
\item two functions $ \sfl_i:\St\to\Val_i$ (lookup) 
\\ and $\sfu_i:\Val_i\times\St\to\St$  (update) 
\item and two equalities
  \begin{equation} 
  \label{eq:states-explicit} 
  \begin{cases} 
  \forall\,a\in \Val_i \,,\; \forall\,s\in \St\,,\;  
     \sfl_i(\sfu_i(a,s)) = a & \\
  \forall\,a\in \Val_i \,,\; \forall\,s\in \St \,,\;   
     \sfl_j(\sfu_i(a,s)) = \sfl_j(s) & \mbox{ for every } j\ne i \in \Loc \\  
  \end{cases} 
  \end{equation} 
\end{itemize} 
Let us assume that $\St=\prod_{i\in\Loc}\Val_i$ 
with the $\sfl_i$'s as projections.
Then two states $s$ and $s'$ are equal 
if and only if $\sfl_i(s)=\sfl_i(s')$ for each $i$, 
and the equalities~\ref{eq:states-explicit}
form a coinductive definition of the functions $\sfu_i$'s. 

\begin{claim}
\label{claim:states-explicit}
This description provides the intended semantics of states. 
\end{claim}

In \cite{PP02} an equational presentation of states is given, 
with seven families of equations. 
In \cite{Me10} these equations are expressed as follows. 
\textit{
\begin{enumerate}
\item 
\textit{Annihilation lookup-update:}
reading the value of
a location $i$ and then updating the location $i$ with the
obtained value is just like doing nothing.
\item 
\textit{Interaction lookup-lookup:}
reading twice the same
location loc is the same as reading it once.
\item 
\textit{Interaction update-update:}
storing a value $a$ and then a value $a'$ 
at the same location~$i$ is just like storing
the value $a'$ in the location. 
\item 
\textit{Interaction update-lookup:}
when one stores a
value $a$ in a location $i$ and then reads the location $i$,
one gets the value $a$. 
\item 
\textit{Commutation lookup-lookup:}
The order of reading
two different locations $i$ and $j$ does not matter.
\item 
\textit{Commutation update-update:}
the order of storing
in two different locations $i$ and $j$ does not matter. 
\item 
\textit{Commutation update-lookup:}
the order of storing
in a location $i$ and reading in another location $j$ does not
matter. 
\end{enumerate}
}
These equations can be translated in our framework as follows, 
with ${\sfl_i}_\modi: \St \to \Val_i\times\St $
defined by ${\sfl_i}_\modi(s)= (\sfl_i(s),s)$
and $\prr_{\Val_i}:\Val_i\times\St \to \St$  
by $\prr_{\Val_i}(a,s)= s$. 
\begin{equation}
\label{eq:states-seven}
\begin{array}{ll}
(1) 
& \forall i\in\Loc ,\; \forall\,s\in\St ,\;\; 
\sfu_i({\sfl_i}_\modi(s)) = s \in\St \\ 
(2) 
& \forall i\in\Loc ,\; \forall\,s\in\St ,\;\;
\sfl_i(\prr_{\Val_i}({\sfl_i}_\modi(s))) = \sfl_i(s) \in\Val_i \\ 
(3) 
& \forall i\in\Loc ,\; \forall\,s\in\St ,\; a,a'\in\Val_i ,\;\; 
\sfu_i(a',\sfu_i(a,s)) = \sfu_i(a',s) \in\St \\ 
(4) 
& \forall i\in\Loc ,\; \forall\,s\in\St ,\; a\in\Val_i ,\;\; 
\sfl_i(\sfu_i(a,s)) = a \in\Val_i \\ 
(5) 
&  \forall i\ne j\in\Loc ,\; \forall\,s\in\St ,\;\; 
   ( \sfl_i(s) , \sfl_j({\sfl_i}_\modi(s)) ) =
   ( \sfl_i({\sfl_j}_\modi(s)) , \sfl_j(s) ) 
   \in\Val_i\times\Val_j \\ 
(6) 
& \forall i\ne j\in\Loc ,\; 
\forall\,s\in\St ,\; a\in\Val_i ,\; b\in\Val_j ,\;\; 
\sfu_j(b,\sfu_i(a,s)) = \sfu_i(a,\sfu_j(b,s)) \in\St \\ 
(7) 
& \forall i\ne j\in\Loc ,\; 
\forall\,s\in\St ,\; a\in\Val_i ,\;\;  
{\sfl_j}_\modi(\sfu_i(a,s)) = (\sfl_j(s), \sfu_i(a,s)) \in \Val_j\times\St \\
\end{array}
\end{equation}

\begin{prop}
\label{proposition:states-explicit-eq}  
Let us assume that $\St=\prod_{i\in\Loc}\Val_i$ 
with the $\sfl_i$'s as projections. 
Then equations~\ref{eq:states-explicit} 
and~\ref{eq:states-seven} are equivalent. 
\end{prop}

In fact, we prove that, without the assumption about $\St$,   
equations~\ref{eq:states-explicit} 
are equivalent to equations~\ref{eq:states-seven} 
considered as \emph{observational} equations: 
two states $s$ and $s'$ are observationaly equivalent  
when $l_k(s)=l_k(s')$ for each location $k$.  
These properties are revisited in proposition~\ref{proposition:states-prop}
and in appendix~\ref{app}. 

\proof
Equations (2) and (5) 
follow immediately from $\prr_{\Val_i}({\sfl_i}_\modi(s))= s$. 
Equation (4) 
is the first equation in~\ref{eq:states-explicit}.
Equation (7) 
is $ (\sfl_j(\sfu_i(a,s)),\sfu_i(a,s)) = (\sfl_j(s), \sfu_i(a,s)) $,
which is equivalent to 
$ \sfl_j(\sfu_i(a,s)) = \sfl_j(s)$:
this is the second equation in~\ref{eq:states-explicit}.
For the remaining equations (1), (3) and (6),  
which return states, it is easy to check that 
by applying $\sfl_k$ to both members 
and using equations~\ref{eq:states-explicit}
we get the same value in $\Val_k$ for each location $k$.  
\qed 

\subsection{Exceptions}
\label{subsec:explicit-exceptions}

The syntax for exceptions heavily depends on the language. 
For instance:
\begin{itemize}
\item In ML-like languages there are several exception names, 
called \emph{constructors}; 
the keywords for raising and handling exceptions are 
\texttt{raise} and \texttt{handle},
which are used in syntactic constructions like: 
\\ \hsp \texttt{ raise i a } and  
\texttt{ ... handle i a => g(a) | j b => h(b) | ... } 
\\ where $i,j$ are exception constructors, 
$a,b$ are parameters and $g,h$ are functions. 
\item In \java there are several exception \emph{types}; 
the keywords for raising and handling exceptions are 
\texttt{throw} and \texttt{try-catch} 
which are used in syntactic constructions like: 
\\ \hsp \texttt{ throw new i(a) } and 
\texttt{ try \{ ... \} catch (i a) {g} catch (j b) {h} ... }
\\ where $i,j$ are exception types, 
$a,b$ are parameters and $g,h$ are programs. 
\end{itemize}

In spite of the differences in the syntax, 
the semantics of exceptions is rather similar in many languages. 
A major point is that there are two kinds of values:
the ordinary (i.e., non-exceptional) values and the exceptions;
it follows that the operations may be classified 
according to the way they may, or may not, interchange 
these two kinds of values. 

First let us focus on the raising of exceptions. 
Let $\Exc$ denote the set of \emph{exceptions}. 
Let $\Const$ denote the set of \emph{exception constructors}. 
For each exception constructor $i$, 
there is a set of \emph{parameters} $\Par_i$ 
and a function $\sft_i:\Par_i\to\Exc$ 
for building the exception $\sft_i(a)$ of constructor $i$ 
with the given parameter $a\in\Par_i$,
called the \emph{key throwing} function.
Then the function $\rais{i}{Y}:\Par_i\to Y+\Exc$
for \emph{raising} (or \emph{throwing}) an exception 
of constructor $i$ into a type $Y$ 
is made of the key throwing function $\sft_i$ followed by 
the inclusion $\inr_Y:\Exc\to Y+\Exc$.
 $$\rais{i}{Y} \;=\; \throw{i}{Y} \;=\;
  \inr_Y \circ \sft_i \;:\; \Par_i \to Y+\Exc  $$
  \begin{equation}
  \label{diag:raise-explicit} 
  \xymatrix@C=4pc@R=1.5pc{
  \Par_i \ar[rr]^{\rais{i}{Y}} \ar@<-.5ex>[rrd]_{\sft_i} && 
    Y+\Exc \ar@{}[lld]|(.3){=} \\ 
  && \Exc \ar[u]_{\inr}  \\ 
  } 
  \end{equation}

\begin{claim}
\label{claim:key-throw}
The function $\sft_i:\Par_i\to\Exc$ is the \emph{key} function 
for \emph{throwing} an exception: 
in the construction of the raising function ($\rais{i}{Y}$), 
only $\sft_i$ turns a non-exceptional value $a\in \Par_i$ 
to an exception $\sft_i(a)\in\Exc$.
\end{claim}

Given a function $f:X\to Y+\Exc$ and an element $x\in X$, 
if $f(x)=\rais{i}{Y}(a)\in Y+\Exc$ for some $a\in\Par_i$
then one says that $f(x)$ \emph{raises an exception of constructor $i$ 
with parameter $a$ into $Y$}.
One says that a function $f:X+\Exc \to Y+\Exc$ \emph{propagates exceptions} 
when it is the identity on $\Exc$. 
Clearly, any function $f:X\to Y+\Exc$ 
can be \emph{extended by propagating exceptions}: 
the extended function $\Ppg(f):X+\Exc \to Y+\Exc$ 
coincides with $f$ on $X$ and with the identity on $\Exc$. 

Now let us study the handling of exceptions, 
starting from its description in \java \cite[Ch.~14]{Java}. 

\textit{
A try statement without a finally block is executed by 
first executing the try block. 
Then there is a choice:
\begin{enumerate}
\item 
\label{tc-one}
If execution of the try block completes normally, 
then no further action is taken 
and the try statement completes normally.
\item 
\label{tc-two}
If execution of the try block completes abruptly 
because of a throw of a value $V$, 
then there is a choice:
  \begin{enumerate}
  \item 
  If the run-time type of $V$ is assignable to the parameter 
  of any catch clause of the try statement, 
  then the first (leftmost) such catch clause is selected. 
  The value $V$ is assigned to the parameter of the selected 
  catch clause, and the block of that catch clause is executed. 
    \begin{enumerate}
    \item
    \label{tc-i}
    If that block completes normally, 
    then the try statement completes normally; 
    \item 
    \label{tc-ii}
    if that block completes abruptly for any reason, 
    then the try statement completes abruptly for the same reason.
    \end{enumerate}
  \item 
  If the run-time type of $V$ is not assignable to the parameter 
  of any catch clause of the try statement, then the try statement 
  completes abruptly because of a throw of the value $V$.
  \end{enumerate}
\item 
\label{tc-three}
If execution of the try block completes abruptly for any other reason, 
then the try statement completes abruptly for the same reason.
\end{enumerate}
}

In fact, points~\ref{tc-i} and~\ref{tc-ii} can be merged. 
Our treatment of exceptions is similar to the one in \java 
when execution of the try block completes normally (point~\ref{tc-one}) 
or completes abruptly because of a throw of an exception of constructor 
$i\in\Const$ (point~\ref{tc-two}).
Thus, for handling exceptions of constructors $i_1,\dots,i_n$ 
raised by some function $f:X\to Y+\Exc$, 
using functions $g_1:\Par_{i_1}\to Y+\Exc,\dots,g_n:\Par_{i_n}\to Y+\Exc$, 
for every $n\geq1$, the handling process builds a function: 
  $$ \handlerec{f}{i_1}{g_1}{i_n}{g_n} \;=\;
  \try{f}{\catchrec{i_1}{g_1}{\catchrec{i_2}{g_2}{...\catch{i_n}{g_n}}}} $$ 
which may be seen, equivalently, either as a function from $X$ to $Y+\Exc$ 
or as a function from $X+\Exc$ to $Y+\Exc$ which propagates the exceptions. 
We choose the second case, and we use compact notations:
  $$ \handlen{f}{(i_k\!\!\Rightarrow\!\! g_k)_{1\leq k\leq n}} \;=\;
    \try{f}{\catchn{i_k\{g_k\}_{1\leq k\leq n}}} : X+\Exc\to Y+\Exc $$
This function can be defined as follows. 
\begin{tabbing}
 9999 \= 9999 \= 9999 \= 9999 \= \kill 
 \> For each $x\in X+\Exc$, 
   $\;(\handlen{f}{(i_k\!\!\Rightarrow\!\! g_k)_{1\leq k\leq n}})(x) \in Y+\Exc$ 
   is defined by: \\
 \>\> if $x\in \Exc$ then return $x\in \Exc \subseteq Y+\Exc$; \\
 \>\> // \textit{now $x$ is not an exception} \\ 
 \>\> compute $y:=f(x) \in Y+\Exc$; \\
 \>\> if $y\in Y$ then return $y\in Y \subseteq Y+\Exc$; \\
 \>\> // \textit{now $y$ is an exception} \\ 
 \>\> for $k=1..n$ repeat \\ 
 \>\>\> if $y=\sft_{i_k}(a)$ for some $a\in \Par_{i_k}$ 
   then return $g_k(a)\in Y+\Exc$; \\
 \>\> // \textit{now $y$ is an exception 
   not constructed from any $i\in\{i_1,\dots,i_n\}$} \\
 \>\> return $y\in \Exc \subseteq Y+\Exc$.  
 \end{tabbing}
In order to express more clearly the apparition of the parameter $a$ 
when $y$ is an exception of constructor $i_k$, 
we introduce for each $i\in\Const$ 
the function $\sfc_i:\Exc\to \Par_i+\Exc$,
called the \emph{key catching} function, defined as follows: 
\begin{tabbing}
 9999 \= 9999 \= \kill 
 \> For each $e\in \Exc$, $\;\sfc_i(e) \in \Par_i+\Exc$ is defined by: \\
 \>\> if $e=\sft_i(a)$ then return $a\in \Par_i \subseteq \Par_i+\Exc$; \\
 \>\> // \textit{now $e$ is an exception not constructed from $i$} \\ 
 \>\> return $e\in \Exc \subseteq \Par_i +\Exc$. 
 \end{tabbing}
This means that the function $\sfc_i$ tests whether the given exception $e$ 
has constructor $i$, 
if so then it \emph{catches the exception} by returning the parameter 
$a\in\Par_i$ such that $e=\sft_i(a)$, 
otherwise $\sfc_i$ propagates the exception $e$. 
Using the key catching function $\sfc_i$, the definition of the handling 
function can be re-stated as follows,
with the three embedded conditionals numerated from the innermost
to the outermost, for future use.  
\begin{tabbing}
 9999 \= 9999 \= 9999 \= 9999 \= \kill 
 \> For each $x\in X+\Exc$, 
   $\;(\handlen{f}{(i_k\!\!\Rightarrow\!\! g_k)_{1\leq k\leq n}})(x) \in Y+\Exc$ 
   is defined by: \\
 \>\> (3) if $x\in \Exc$ then return $x\in \Exc \subseteq Y+\Exc$; \\
 \>\> // \textit{now $x$ is not an exception} \\ 
 \>\> compute $y:=f(x) \in Y+\Exc$; \\
 \>\> (2) if $y\in Y$ then return $y\in Y \subseteq Y+\Exc$; \\
 \>\> // \textit{now $y$ is an exception} \\ 
 \>\> for $k=1..n$ repeat \\ 
 \>\>\> compute $y:=\sfc_{i_k}(y)\in \Par_{i_k}+\Exc$;  \\
 \>\>\> (1) if $y\in \Par_{i_k}$ then return $g_k(y)\in Y+\Exc$; \\
 \>\> // \textit{now $y$ is an exception 
    not constructed from any $i\in\{i_1,\dots,i_n\}$} \\
 \>\> return $y\in \Exc \subseteq Y+\Exc$. 
 \end{tabbing}
Note that whenever several $i$'s are equal in $(i_1,\dots,i_n)$, 
then only the first $g_i$ may be used.

\begin{claim}
\label{claim:key-catch}
The function $\sfc_i:\Exc\to \Par_i+\Exc$ is the \emph{key} function 
for \emph{catching} an exception: 
in the construction of the handling function ($\handle{f}{i}{g}$), 
only $\sfc_i$ may turn an exception $e\in\Exc$ 
to a non-exceptional value $\sfc_i(e)\in\Par_i$,
the other parts of the construction propagate all exceptions. 
\end{claim}

The definition of the handling function is 
illustrated by the following diagrams;
each diagram corresponds to one of the three nested conditionals,
from the innermost to the outermost.
The inclusions are denoted by 
$\inl_A : A \to A+\Exc $ and $\inr_A : \Exc \to A+\Exc $
(subscripts may be dropped) and for every $a:A\to B$ and $e:\Exc\to B$
the corresponding conditional is denoted by $\cotuple{a\,|\,e}:A+\Exc\to B$,
it is characterized by the equalities 
$ \cotuple{a\,|\,e} \circ \inl_A  = a$ and 
$ \cotuple{a\,|\,e} \circ \inr_A  = e$.
\begin{enumerate}
\item The catching functions $\catchn{i_k\{g_k\}_{p\leq k\leq n}} : \Exc\to Y+\Exc$
are defined recursively by 
  $$ \catchn{i_k\{g_k\}_{p\leq k\leq n}} \;=\;
    \begin{cases} 
     \cotuple{ g_n \;|\; \inr_Y  } \circ \sfc_{i_n} & \mbox{ when } p=n \\
     \cotuple{ g_p \;|\; \catchn{i_k\{g_k\}_{p+1\leq k\leq n}} } \circ \sfc_{i_p} 
       & \mbox{ when } p<n \\
   \end{cases} $$ 
  \begin{equation}
  \label{diag:handle-explicit-catch} 
  \xymatrix@C=3pc{
  & \Par_{i_p} \ar[d]_{\inl} \ar[rrrrd]^{g_p} &&&& \\
  \Exc \ar[r]^(.4){\sfc_{i_p}} & 
    \Par_{i_p}+\Exc 
      \ar[rrrr]^(.4){\cotuple{ g_p \;|\; \dots } } &
    \ar@{}[ul]|(.4){=} \ar@{}[dl]|(.4){=} &&& Y+\Exc \\ 
  & \Exc \ar[u]^{\inr} \ar[rrrru]_{\dots} &&&& \\
  }  
  \end{equation}
where $\dots$ stands for $\inr_Y$ when $p=n$ 
and for $\catchn{i_k\{g_k\}_{p+1\leq k\leq n}} $ when $p<n$.
\item Then the function $\Handle:X\to Y+\Exc$,
which defines the handling function on non-exceptional values, 
is defined as 
$$  \Handle \;=\; 
    \cotuple{ \inl_Y \;|\; \catchn{i_k\{g_k\}_{1\leq k\leq n} }} \circ f : 
    X\to Y+\Exc $$
  \begin{equation}
  \label{diag:handle-explicit-intermediate} 
  \xymatrix@C=3pc{
  & Y \ar[d]_{\inl} \ar[rrrrd]^{\inl} &&&& \\
  X \ar[r]^{f} & 
    Y+\Exc \ar[rrrr]^(.4){\cotuple{ \inl \;|\; \catchn{i_k\{g_k\}_{1\leq k\leq n} }}} &
    \ar@{}[ul]|(.4){=} \ar@{}[dl]|(.4){=} &&& Y+\Exc \\ 
  & \Exc \ar[u]^{\inr} \ar[rrrru]_{\catchn{i_k\{g_k\}_{1\leq k\leq n}}} &&&& \\
  } 
  \end{equation}  
\item Finally the handling function is the extension of $\Handle$
which propagates exceptions 
$$ \try{f}{\catchn{i_k\{g_k\}_{1\leq k\leq n}}} \;=\; 
  \cotuple{ \Handle \;|\; \inr_Y } $$
  \begin{equation}
  \label{diag:handle-explicit-handle} 
  \xymatrix@C=3pc{
  X \ar[d]_{\inl} \ar[rrrrd]^{\Handle} &&&& \\
  X+\Exc \ar[rrrr]^(.4){\try{f}{\catchn{i_k\{g_k\}_{1\leq k\leq n}}}} &
    \ar@{}[ul]|(.4){=} \ar@{}[dl]|(.4){=} &&& Y+\Exc \\ 
  \Exc \ar[u]^{\inr} \ar[rrrru]_{\inr} &&&& \\
  } 
  \end{equation}
\end{enumerate}

The next claim is based on our previous analysis of \java exceptions; 
it is also related to the notion of \emph {monadic reflection} in 
\cite{Fi94}.

\begin{claim}
\label{claim:exceptions-explicit}
This description provides the intended semantics of exceptions. 
\end{claim}

Let us come back to the key operations $\sft_i$ and $\sfc_i$
for throwing and catching exceptions. 
For each $i\in\Const$ there are:
\begin{itemize}
\item a set $\Par_i$ (parameters)
\item two functions $\sft_i:\Par_i\to\Exc$ (key throwing)
\\ and $\sfc_i:\Exc\to\Par_i+\Exc$ (key catching)
\item and two equalities
  \begin{equation} 
  \label{eq:exceptions-explicit} 
  \begin{cases} 
  \forall\,a\in \Par_i \,,\;
    \sfc_i(\sft_i(a)) = a\in \Par_i \subseteq \Par_i+\Exc & \\
  \forall\,b\in \Par_j \,,\;
    \sfc_i(\sft_j(b)) = \sft_j(b)\in \Exc \subseteq \Par_i+\Exc
    & \mbox{ for every } j\ne i \in \Loc \\  
  \end{cases} 
  \end{equation} 
\end{itemize} 
This means that, given an exception $e$ of the form $t_i(a)$, 
the corresponding key catcher $\sfc_i$ recovers the non-exceptional value $a$
while the other key catchers propagate the exception~$e$. 
Let us assume that $\Exc=\sum_{i\in\Const}\Par_i$ 
with the $\sft_i$'s as coprojections.
Then the equalities ~\ref{eq:exceptions-explicit}
form an inductive definition of the functions $\sfc_i$'s. 

\subsection{States and exceptions: the duality}
\label{subsec:explicit-symmetry}

Figure~\ref{fig:duality-semantics} recapitulates the properties of 
the functions \emph{lookup} ($\sfl_i$) and \emph{update} ($\sfu_i$) 
for states on the left, 
and the functions \emph{key throw} ($\sft_i$) and 
\emph{key catch} ($\sfc_i$) for exceptions on the right. 
Intuitively:
for looking up the value of a location $i$,
only the \emph{previous} updating of this location is necessary, 
and dually, 
when throwing an exception of constructor $i$
only the \emph{next} catcher for this constructor is necessary 
(see section~\ref{subsec:dual-other}). 
The next result follows immediately from figure~\ref{fig:duality-semantics}. 

\begin{figure}[!h]
\renewcommand{\arraystretch}{1.3}
$$ \begin{array}{|c|c|}
\hline
\makebox[60mm]{\textbf{States}} & 
\makebox[60mm]{\textbf{Exceptions}} \\ 
\hline
i\in\Loc,\; \Val_i,\; &
i\in\Const,\; \Par_i,\; \\
\St\,(=\prod_{i\in\Loc}\Val_i) &
\Exc\,(=\sum_{i\in\Const}\Par_i) \\
\mbox{cartesian products:} & \mbox{disjoint unions:} \\
\xymatrix@C=1.5pc{
\Val_i & \Val_i\times\St \ar[l]_(.6){\prl_i} \ar[r]^(.6){\prr_i} & \St \\ 
} & 
\xymatrix@C=1.5pc{
\Exc \ar[r]^(.4){\inr_i} & \Par_i+\Exc & \Par_i \ar[l]_(.4){\inl_i} \\ 
} \\
\hline
\sfl_i:\St\to\Val_i &
\Exc\lto\Par_i:\sft_i \\ 
\sfu_i:\Val_i\times\St\to\St &
\Par_i+\Exc\lto\Exc:\sfc_i \\ 
\hline
\xymatrix@R=1pc@C=3pc{
\Val_i\times\St \ar[r]^{\prl_i} \ar[d]_{\sfu_i} & \Val_i \ar[d]^{\id} \\
\St \ar[r]^{\sfl_i} & \Val_i \ar@{}[ul]|{=} \\
} &
\xymatrix@R=1pc@C=3pc{
\Par_i+\Exc & \Par_i \ar[l]_{\inl_i} \\
\Exc \ar[u]^{\sfc_i} & \Par_i \ar[l]_{\sft_i} \ar[u]_{\id} \ar@{}[ul]|{=} \\
} \\
\xymatrix@R=1pc@C=2pc{
\Val_i\times\St \ar[r]^{\prr_i} \ar[d]_{\sfu_i} & 
  \St \ar[r]^{\sfl_j} & \Val_j \ar[d]^{\id} \\
\St \ar[rr]^{\sfl_j} && \Val_j \ar@{}[ull]|{=} \\
} & 
\xymatrix@R=1pc@C=2pc{
\Par_i+\Exc & \Exc \ar[l]_{\inr_i}
  & \Par_j \ar[l]_{\sft_j} \\
\Exc \ar[u]^{\sfc_i} && \Par_j \ar[ll]_{\sft_j} \ar[u]_{\id} \ar@{}[ull]|{=} \\
}  \\
(j\ne i) & (j\ne i) \\ 
\hline
\end{array} $$
\renewcommand{\arraystretch}{1}
\caption{Duality of semantics}
\label{fig:duality-semantics}
\end{figure}

\begin{prop}
\label{proposition:duality-explicit}
The well-known duality between categorical products and coproducts 
can be extended as a duality 
between the semantics of the lookup and update functions 
for states on one side 
and the semantics of the key throwing and catching functions 
for exceptions on the other.
\end{prop}

It would be unfair to consider states and exceptions only from this
denotational point of view. Indeed, states and exceptions are
\emph{computational effects}, which do not appear explicitly in the syntax:
in an imperative language there is no type of states, 
and in a language with exceptions 
the type of exceptions that may be raised by a program 
is not seen as a return type for this program. 
In fact, our result (theorem~\ref{theorem:duality}) 
is that there is a duality between states and exceptions 
considered as computational effects, 
which provides the above duality 
(propostion~\ref{proposition:duality-explicit}) 
between their semantics.

\section{Computational effects} 
\label{sec:effects}

In sections~\ref{sec:states} and~\ref{sec:exceptions} 
we will deal with states and exceptions as computational effects.
In this section, we present our point of view on computational effects. 
First a motivating example from object-oriented programming is given, 
then a simplified version of the framework of diagrammatic logics 
is presented, 
and finally this framework is applied to effects. 

\subsection{An example} 
\label{subsec:bank}

In this section we use a toy example 
dealing with the state of an object in an object-oriented language, 
in order to outline our approach of computational effects. 
Let us build a class \texttt{BankAccount} for managing (very simple!) 
bank accounts. We use the types \texttt{int} and \texttt{void}, 
and we assume that \texttt{int} is interpreted as the set of integers $\bZ$ 
and \texttt{void} as a singleton $\bU$. 
In the class \texttt{BankAccount}, there is a method 
\texttt{balance()} which returns the current balance of the account
and a method \texttt{deposit(x)} for the deposit 
of \texttt{x} Euros on the account. 
The \texttt{deposit} method is a \emph{modifier},
which means that it can use and modify the state of the current account.
The \texttt{balance} method 
is an \emph{inspector}, or an \emph{accessor}, which means that 
it can use the state of the current account but it is not allowed to modify 
this state.
In the object-oriented language \cpp,
a method is called a \emph{member function};
by default a member function is a modifier, 
when it is an accessor it is called a \emph{constant member function}
and the keyword \texttt{const} is used.
So, the \cpp syntax for declaring the member functions
of the class \texttt{BankAccount} looks like:
\begin{center}
\begin{tabular}{l}
  \texttt{int} \texttt{balance} (\,)  \texttt{const} ; \\
  \texttt{void} \texttt{deposit} (\texttt{int}) ; \\
\end{tabular}
\end{center}

Forgetting the keyword \texttt{const}, 
this piece of \cpp syntax can be translated as a signature $\Ss_{\bank,\app}$,
which we call the \emph{apparent signature}:
\begin{equation}
\label{equation:bank-app}
  \Ss_{\bank,\app}: \begin{cases}
  \texttt{balance}:\texttt{void}\to\texttt{int} \cr
   \texttt{deposit}:\texttt{int}\to\texttt{void} 
 \end{cases} 
\end{equation}
In a model (or algebra) of the signature $\Ss_{\bank,\app}$, 
the operations would be interpreted as functions:
$$
  \begin{cases}
  \deno{\texttt{balance}}:\bU\to\bZ \cr
  \deno{\texttt{deposit}}:\bZ\to\bU 
 \end{cases} 
$$
which clearly is not the intended interpretation.

In order to get the right semantics, 
we may use another signature $\Ss_{\bank,\expl}$, 
which we call the \emph{explicit signature},
with a new symbol \texttt{state} for the ``type of states'':
\begin{equation}
\label{equation:bank-expl}
  \Ss_{\bank,\expl}: \begin{cases}
  \texttt{balance}:\texttt{state}\to\texttt{int} \cr
   \texttt{deposit}:\texttt{int}\times\texttt{state}\to\texttt{state} 
 \end{cases} 
\end{equation}
The intended interpretation is a model of the explicit signature 
$\Ss_{\bank,\expl}$, 
with $\St$ denoting the set of states of a bank account:
$$
  \begin{cases}
  \deno{\texttt{balance}}:\St\to\bZ \cr
  \deno{\texttt{deposit}}:\bZ\times\St\to\St 
 \end{cases} 
$$

So far, in this example, we have considered two different signatures.
On the one hand, the apparent signature $\Ss_{\bank,\app}$
is simple and quite close to the \cpp code, but 
the intended semantics is not a model of  $\Ss_{\bank,\app}$.
On the other hand, the semantics is a model of the explicit 
signature $\Ss_{\bank,\expl}$, 
but $\Ss_{\bank,\expl}$ is far from the \cpp syntax: actually, the very nature of 
the object-oriented language is lost by introducing a ``type of states''. 
Let us now define a \emph{decorated signature} $\Ss_{\bank,\deco}$,
which is still closer to the \cpp code than the apparent signature 
and which has a model corresponding to the intended semantics.
The decorated signature is not exactly a signature in the classical sense, 
because there is a classification of its operations.
This classification is provided by superscripts called \emph{decorations}: 
the decorations ``\texttt{(1)}'' and ``\texttt{(2)}'' correspond respectively 
to the object-oriented notions of \emph{accessor} and \emph{modifier}.  
\begin{equation}
\label{equation:bank-deco}
  \Ss_{\bank,\deco}:   \begin{cases}
  \texttt{balance}^\texttt{(1)}:\texttt{void}\to\texttt{int} \cr
  \texttt{deposit}^\texttt{(2)}:\texttt{int}\to\texttt{void} 
  \end{cases} 
\end{equation}
The decorated signature is similar to the \cpp code,
with the decoration ``\texttt{(1)}'' corresponding to the keyword ``\texttt{const}''.
In addition, we claim that the intended semantics 
can be seen as a \emph{decorated model} of this decorated signature.

In order to add to the signature the constants of type \texttt{int} 
like \texttt{0}, \texttt{1}, \texttt{2}, \dots and the usual operations on integers, 
a third decoration is used: 
the decoration ``\texttt{(0)}'' for \emph{pure} functions, 
which means, for functions which neither inspect nor modify 
the state of the bank account. 
So, we add to the apparent and explicit signatures the constants 
$\texttt{0},\;\texttt{1},\;$\dots$:\texttt{void}\to\texttt{int}$ 
and the operations $\texttt{+},\;\texttt{-},\;\texttt{$\ast$}: 
\texttt{int}\times\texttt{int} \to\texttt{int}$, 
and we add to the decorated signature the pure constants 
$\texttt{0}^\texttt{(0)},\;\texttt{1}^\texttt{(0)},\;$\dots$:
\texttt{void}\to\texttt{int}$
and the pure operations 
$\texttt{+}^\texttt{(0)},\;\texttt{-}^\texttt{(0)},\texttt{$\ast$}^\texttt{(0)}: 
\texttt{int}\times\texttt{int} \to\texttt{int}$.
For instance in the \cpp expressions  
  $$ \texttt{deposit(7); balance()} \;\mbox{ and }\; \texttt{7 + balance()}$$ 
composition is expressed in several different ways:
in the functional way $f(a)$,
in the infix way $a\,f\,b$
and in the imperative way $c;c'$. 
In the explicit signature, these expressions can be seen as the terms 
$\texttt{balance}\circ \texttt{deposit}\circ 
(\texttt{7}\times\texttt{id}_\texttt{state})$ 
and $\texttt{+}\circ(\texttt{7} \times \texttt{balance})$, 
with $\texttt{void}\times\texttt{state}$ identified with $\texttt{state}$: 
  $$ \xymatrix@C=4pc@R=.5pc{ 
  \texttt{state} \simeq
  \texttt{void}\times\texttt{state} 
    \ar[r]^(.6){\texttt{7}\times\texttt{id}_\texttt{state}} & 
  \texttt{int}\times\texttt{state} \ar[r]^(.6){\texttt{deposit}} & 
  \texttt{state} \ar[r]^{\texttt{balance}} & 
  \texttt{int} \\
  \texttt{state} \simeq
  \texttt{void}\times\texttt{state} 
    \ar[r]^(.6){\texttt{7}\times\texttt{balance}} & 
  \texttt{int}\times\texttt{int} \ar[r]^(.6){\texttt{+}} & 
  \texttt{int} \\
  }$$
In the decorated signature, they can be seen as the decorated terms
\\ $\texttt{balance}^\texttt{(1)}\circ \texttt{deposit}^\texttt{(2)}\circ 
  \texttt{7}^\texttt{(0)}$ and 
$ \texttt{+}^\texttt{(0)}\circ \tuple{\texttt{7}^\texttt{(0)},\texttt{balance}^\texttt{(1)}}$:
  $$ \xymatrix@C=6pc@R=.5pc{
  \texttt{void} \ar[r]^{\texttt{7}^\texttt{(0)}} & 
  \texttt{int} \ar[r]^{\texttt{deposit}^\texttt{(2)}} & 
  \texttt{void} \ar[r]^{\texttt{balance}^\texttt{(1)}} & 
  \texttt{int} \\
  \texttt{void} \ar[r]^{\tuple{\texttt{7}^\texttt{(0)},\texttt{balance}^\texttt{(1)}}} & 
  \texttt{int}\times\texttt{int} \ar[r]^{\texttt{+}^\texttt{(0)}} & 
  \texttt{int} \\
  }$$
These two expressions have different effects: 
the first one is a modifier while the second one is an accessor;
however, both return the same result (an integer).
We introduce the symbol $\eqw$ for the relation ``same result,
maybe distinct effects''; 
the relation $\eqw$ will be considered as a decorated version of the equality. 
$$\texttt{balance}^\texttt{(1)}\circ \texttt{deposit}^\texttt{(2)}\circ 
  \texttt{7}^\texttt{(0)}
\;\eqw\; 
\texttt{+}^\texttt{(0)}\circ\tuple{\texttt{7}^\texttt{(0)},\texttt{balance}^\texttt{(1)}}$$

\subsection{Simplified diagrammatic logics}
\label{subsec:dialog} 

In this paper, as in~\cite{DD10} and \cite{DDR11},
we use the point of view of \emph{diagrammatic logics}
for dealing with computational effects.
One fundamental feature of the theory of
diagrammatic logics is the distinction 
between a logical theory and its presentations
(or specifications). 
This is the usual point of view in the framework 
of algebraic specifications \cite{EM85},
but not always in logic, as mentioned  
by F.W. Lawvere in his foreword to \cite{ARV11}:
\textit{
Yet many works in general algebra (and model theory generally) 
continue anachronistically to confuse 
a presentation in terms of signatures with the presented theory itself.
}
A second fundamental feature of the theory of
diagrammatic logics is the 
definition of a rich family of morphisms of logics.
Computational effects, from our point of view,
heavily depend on some morphisms of logics. 
Thus, in this paper, 
in order to focus on states and exceptions as effects, 
we use a simplified version of diagrammatic logics 
by dropping the distinction between 
a logical theory and its presentations. 
It is only in remark~\ref{remark:dialog-dialog} that we give 
some hints about non-simplified diagrammatic logics. 

On the other hand, with the same goal  
of focusing on states and exceptions as effects, 
in sections~\ref{sec:states} and~\ref{sec:exceptions} 
the base logic is the very simple  
(multi-sorted) \emph{monadic equational logic}, 
where a theory is made of types, unary terms and equations. 
We will occasionally mention the \emph{equational logic}, 
where in addition a theory may have terms of any finite arity.  
In order to keep the syntactic aspect of the logics, 
we use a congruence relation between terms 
rather than the equality;
in the denotational semantics, this congruence is usually 
interpreted as the equality. 

\begin{defi}
\label{definition:dialog-logic}
A \emph{simplified diagrammatic logic} is a category $\catT$ with colimits; 
its objects are called the $\catT$-\emph{theories}
and its morphisms the \emph{morphisms of $\catT$-theories}.
A \emph{morphism of simplified diagrammatic logics} $F:\catT\to\catT'$
is a left adjoint functor. 
This yields the \emph{category of simplified diagrammatic logics}. 
\end{defi}

\begin{exa}[Monadic equational logic]
\label{example:dialog-meqn}
A monadic equational theory might be called a ``syntactic category'':
it is a category where the axioms hold only up to some congruence relation.
Precisely, a \emph{monadic equational theory} is a directed graph 
(its vertices are called \emph{objects} or \emph{types} 
and its edges are called \emph{morphisms} or \emph{terms})
with an \emph{identity} term $\id_X:X\to X$ for each type $X$ 
and a \emph{composed} term $g\circ f:X\to Z$ for each pair 
of consecutive terms $(f:X\to Y,g:Y\to Z)$;
in addition it is endowed with \emph{equations} $f\eqs g:X\to Y$
that form an equivalence relation on parallel terms, denoted by $\eqs$,
which is a \emph{congruence} with respect to the composition 
and such that the associativity and identity axioms hold up to 
congruence.
This definition of the monadic equational logic 
can be described by a set of \emph{inference rules},
as in figure~\ref{fig:meqn-rules}.  
A morphism of monadic equational theories might be called 
a ``syntactic functor'':
it maps types to types, terms to terms and equations to equations. 

\begin{figure}[!h]
\renewcommand{\arraystretch}{3}
$$ \begin{array}{|c|} 
\hline 
\rncomp \dfrac{f:X\to Y \quad g:Y\to Z}{g\circ f:X\to Z} \qquad 
\rnid \dfrac{X}{\id_X:X\to X } \\ 
\rnassoc \dfrac{f:X\to Y \quad g:Y\to Z \quad h:Z\to W}
  {h\circ (g\circ f) \eqs (h\circ g)\circ f} \\
\rnidsrc \dfrac{f:X\to Y}{f\circ \id_X \eqs f} \qquad 
\rnidtgt \dfrac{f:X\to Y}{\id_Y\circ f \eqs f} \\ 
\rnsrefl \dfrac{}{f \eqs f} \qquad 
\rnssym \dfrac{f \eqs g}{g \eqs f} \qquad 
\rnstrans\dfrac{f \eqs g \quad g \eqs h}{f \eqs h} \\ 
\rnssubs \dfrac{f:X\to Y \quad g_1\eqs g_2:Y\to Z}
  {g_1\circ f \eqs g_2\circ f :X\to Z}  \\
\rnsrepl \dfrac{f_1\eqs f_2:X\to Y \quad g:Y\to Z}
  {g\circ f_1 \eqs g\circ f_2 :X\to Z} 
\rule[-20pt]{0pt}{0pt} \\
\hline 
\end{array}$$
\renewcommand{\arraystretch}{1}
\caption{Rules of the monadic equational logic}
\label{fig:meqn-rules}
\end{figure}
\end{exa}

\begin{exa}[Equational logic]
\label{example:dialog-eqn}
An equational theory might be called 
a ``syntactic category with finite products''.
Precisely, an \emph{equational theory} is a monadic equational theory 
with in addition, for each finite family $(Y_i)_{1\leq i\leq n}$ of types, 
a \emph{product (up to congruence)} made of a cone 
$(q_i:\prod_{j=1}^nY_j \to Y_i)_{1\leq i\leq n} $ 
such that for each cone $(f_i:X \to Y_i)_{1\leq i\leq n} $
with the same base 
there is a term $\tuple{f_1,\dots,f_n}:X\to \prod_{j=1}^nY_j$
such that $q_i\circ \tuple{f_1,\dots,f_n} \eqs f_i$ for each $i$,
and whenever some $g:X\to \prod_{j=1}^nY_j$ is such that  
$q_i\circ g \eqs f_i$ for each $i$
then $g\eqs \tuple{f_1,\dots,f_n}$.
When $n=0$ this means that in an equational theory 
there is a \emph{terminal} type
$\unit$ such that for each type $X$
there is a term $\tu_X:X\to \unit$,
which is unique up to congruence in the sense that 
every $g:X\to \unit$ satisfies $g\eqs \tu_X$.
A morphism of equational theories is a morphism of monadic equational theories 
which preserves products. 
This definition can be described by a set of \emph{inference rules},
as in figure~\ref{fig:eqn-rules}.  
When there are several parts in the conclusion of a rule,
this must be understood as a
conjunction (which might be avoided by writing several rules). 
The monadic equational logic may be seen as the restriction of 
the equational logic to terms with exactly one ``variable''. 
The functor which maps each monadic equational theory
to its generated equational theory is a morphism of simplified 
diagrammatic logics, with right adjoint the forgetful functor.

\begin{figure}[!h]
\renewcommand{\arraystretch}{3}
$$ \begin{array}{|c|c|} 
\hline 
\mbox{ Rules of the monadic equational logic, and for each $n\in\bN$: } &
\mbox{ i.e., when $n=0$: } \\ 
\dfrac{Y_1\;\dots\; Y_n}{(q_i: \prod_{j=1}^n Y_j \to Y_i)_{1\leq i \leq n}} &   
\dfrac{}{\unit} \\  
\dfrac{(q_i: \prod_{j=1}^n Y_j \to Y_i)_{1\leq i \leq n} \quad 
  (f_i: X \to Y_i)_{1\leq i \leq n}}
  {\tuple{f_1,\dots,f_n}:X\to \prod_{j=1}^n Y_j \quad 
  \forall i \; q_i\circ\tuple{f_1,\dots,f_n} \eqs f_i } & 
\dfrac{X}{\tu_X:X\to \unit } \\ 
  \dfrac{(q_i: \prod_{j=1}^n Y_j \to Y_i)_{1\leq i \leq n} \quad 
  g:X\to \prod_{j=1}^n Y_j \quad 
  \forall i\; q_i \circ g \eqs f_i}
  {g \eqs \tuple{f_1,\dots,f_n}} & 
  \dfrac{g:X\to \unit}{g \eqs \tu_X} 
\rule[-20pt]{0pt}{0pt} \\
\hline 
\end{array}$$
\renewcommand{\arraystretch}{1}
\caption{Rules of the equational logic}
\label{fig:eqn-rules}
\end{figure}
\end{exa}

Given a simplified diagrammatic logic, 
we define the associated notions of model and inference system. 
We often write ``logic'' instead of ``simplified diagrammatic logic''. 

\begin{defi}
\label{definition:dialog-model}
Let $\catT$ be a logic.
Let $\Ff$ and $\Tt$ be $\catT$-theories,
a \emph{model} of $\Ff$ in $\Tt$ is a morphism from $\Ff$ to $\Tt$ in $\catT$. 
Then the triple $\Lang=(\Ff,\Tt,M)$ is a \emph{language} on $\catT$
with \emph{syntax} $\Ff$ and \emph{semantics} $M$.
The set of models of $\Ff$ in $\Tt$ is denoted by $\Mod_{\catT}(\Ff,\Tt)$. 
\end{defi}

\begin{rem}
\label{remark:dialog-sound} 
The definitions are such that 
every simplified diagrammatic logic $\catT$ has the \emph{soundness} property: 
in every language, the semantics is a model of the syntax. 
\end{rem}

\begin{defi}
\label{definition:dialog-rule}
Let $\catT$ be a logic.
An \emph{inference rule} is a morphism $\rho:\Cc\to\Hh$ in $\catT$.
Then $\Hh$ is the \emph{hypothesis} and $\Cc$ is the \emph{conclusion} 
of the rule $\rho$. 
Let $\Ff_0$ and $\Ff$ be $\catT$-theories,
an \emph{instance} of $\Ff_0$ in $\Ff$
is a morphism $\kappa:\Ff_0\to\Ff$ in $\catT$. 
The \emph{inference step} applying a rule $\rho:\Cc\to\Hh$
to an instance $\kappa:\Hh\to\Ff$ of $\Hh$ in $\Ff$ 
is the composition in $\catT$, which builds the instance 
$\kappa\circ\rho:\Cc\to\Ff$ of $\Cc$ in $\Ff$.
\end{defi}

\begin{rem}
The rule $\rho:\Cc\to\Hh$ may be represented 
in the usual way as a ``fraction'' $\frac{\Hh}{\rho(\Cc)}$, 
or as $\frac{\Hh_1,\dots,\Hh_k}{\rho(\Cc)}$ 
when $\Hh$ is the colimit of several theories, 
see example~\ref{example:dialog-rule}.
In addition, in \cite{DD10} it is explained why an inference rule 
written in the usual way as a ``fraction'' $\frac{\Hh}{\rho(\Cc)}$ 
is really a \emph{fraction} in the categorical sense of \cite{GZ67}, 
but with $\Hh$ on the denominator side and $\Cc$ on the numerator side.
\end{rem}

\begin{exa}[Composition rule]
\label{example:dialog-rule}
Let us consider the equational logic $\catT_{\eqn}$,
as in example~\ref{example:dialog-eqn}.
The category of sets can be seen as an equational theory $\Tt_{\set}$,
with the equalities as equations and the cartesian products as products. 
Let us define the equational theory ``of integers'' $\Ff_{\inte}$
as the equational theory 
generated by a type $I$, 
three terms $z:\unit\to I$ and $s,p:I \to I$  
and two equations $s\circ p \eqs \id_I$ and $p\circ s \eqs \id_I$. 
Then there is a unique model $M_{\inte}$ of $\Ff_{\inte}$ in $\Tt_{\set}$ 
which interprets 
the sort $I$ as the set $\bZ$ of integers,
the constant term $z$ as $0$ and the terms $s$ and $p$ as the 
functions $x\mapsto x+1$ and $x\mapsto x-1$. 
In the equational logic $\catT_{\eqn}$,
let us consider the composition rule:
  $$ \frac{f:X\to Y \qquad g:Y\to Z}{g\circ f:X\to Z} $$ 
Let $\Hh$ be the equational theory generated by three
types $X$, $Y$, $Z$ and two consecutive terms $f:X\to Y$, $g:Y\to Z$;
let $\Cc$ be the equational theory generated by two types $T$, $T'$
and a term $t:T\to T'$.
The composition rule corresponds to the morphism of equational theories 
from $\Cc$ to $\Hh$ which maps $t$ to $g\circ f$. 
Let us consider the instance $\kappa$ of $\Hh$ in $\Ff_{\inte}$ 
which maps $f$ and $g$ respectively to $z$ and $s$, 
then the inference step applying the composition rule to this instance $\kappa$
builds the instance of $\Cc$ in $\Ff_{\inte}$ which maps $t$ to $s\circ z$,
as required.
Moreover, $\Hh$ can be obtained as the pushout of 
$\Hh_1$ (generated by $X$, $Y$ and $f:X\to Y$)
and $\Hh_2$ (generated by $Y$, $Z$ and $g:Y\to Z$) 
on their common part (the equational theory generated by $Y$). 
Then the instance $\kappa$ of $\Hh$ in $\Ff_{\inte}$ can be built from
the instance $\kappa_1$ of $\Hh_1$ in $\Ff_{\inte}$
mapping $f$ to $z$ 
and the instance $\kappa_2$ of $\Hh_2$ in $\Ff_{\inte}$
mapping $g$ to $s$. 
\end{exa}

\begin{rem}
\label{remark:dialog-dialog}
In this simplified version of diagrammatic logic,
the morphisms of theories serve for many purposes. 
However in the non-simplified version there is a distinction between 
theories and their presentations (called \emph{specifications}), 
which results in more subtle definitions.
This is outlined here, more details can be found in \cite{DD10}.
This will not be used in the next sections. 
As usual a \emph{locally presentable category} is a category $\catC$ 
which is equivalent to the category of set-valued realizations 
(or models) of a limit sketch \cite{GU71}.
In addition, a functor $F\colon \catC_1\to\catC_2$ which is 
the left adjoint to the precomposition with some morphism 
of limit sketches \cite{Eh68}
will be called a \emph{locally presentable functor}. 
\begin{itemize}
\item A \emph{diagrammatic logic} is defined as a locally presentable functor 
$L:\catS\to\catT$
such that its right adjoint $R$ is full and faithful.
This means that $L$ is a \emph{localization}, 
up to an equivalence of categories: 
it consists of adding inverse morphisms for some morphisms, 
constraining them to become isomorphisms \cite{GZ67}.
The categories $\catS$ and $\catT$ are called the category of \emph{specifications}
and the category of \emph{theories}, respectively, 
of the diagrammatic logic $L$. 
A specification $\Ss$ \emph{presents} a theory $\Tt$
if $\Tt$ is isomorphic to $L(\Ss)$.
The fact that $R$ is full and faithful means that every theory $\Tt$, 
when seen as a specification $R(\Tt)$, presents itself. 
\item A \emph{model} $M$ of a specification $\Ss$ in a theory $\Tt$ is 
a morphism of theories $M\colon L\Ss \to \Tt$
or equivalently, thanks to the adjunction, 
a morphism of specifications $M\colon \Ss \to R\Tt$. 
\item An \emph{entailment} is a morphism $\tau$ in $\catS$ such that 
$L\tau$ is invertible in $\catT$; 
a similar notion can be found in \cite{Ma97}. 
An \emph{instance} $\kappa$ of a specification $\Ss_0$ 
in a specification $\Ss$
is a cospan in $\catS$ made of a morphism $\ss:\Ss_0\to\Ss'$
and an entailment $\tau:\Ss\to\Ss'$. 
It is also called a \emph{fraction} with \emph{numerator} $\ss$ and 
\emph{denominator} $\tau$ \cite{GZ67}. 
The instances can be composed in the usual way as cospans, 
thanks to pushouts in $\catS$.
This forms the \emph{bicategory of instances} of the logic, 
and $\catT$ is, up to equivalence, 
the quotient category of this bicategory.  
An \emph{inference rule} $\rho$ with \emph{hypothesis} $\Hh$ 
and \emph{conclusion} $\Cc$ is an instance of $\Cc$ in $\Hh$. 
Then an inference step is a composition of fractions.
\item An \emph{inference system} for a diagrammatic logic $L$
is a morphism of limit sketches which gives rise to  
the locally presentable functor $L$. 
The \emph{elementary} inference rules are the rules 
in the image of the inference system
by the Yoneda contravariant functor.
Then a \emph{derivation}, or \emph{proof}, 
is the description of a fraction in terms of elementary inference rules.
\item A \emph{morphism of logics} $F\colon L_1\to L_2$,
where $L_1:\catS_1\to\catT_1$ and $ L_2:\catS_2\to\catT_2$,  
is a pair of locally presentable functors 
$(F_S,F_T)$ with $F_S:\catS_1\to\catS_2$ and $F_T:\catT_1\to\catT_2$, 
together with a natural isomorphism $F_T\circ L_1 \iso L_2\circ F_S$ 
induced by a commutative square of limit sketches.
\end{itemize}
\end{rem}

\subsection{Diagrammatic logics for effects}
\label{subsec:effects} 

Now let us come back to computational effects. 
Our point of view is that a language with computational effect is 
a kind of language with an \emph{apparent lack of soundness}: 
a language with computational effect is made
of a syntax, called the \emph{apparent syntax},
and a semantics which (in general) \emph{is not} 
a model of the apparent syntax, 
together with some additional information which may be added 
to the apparent syntax 
in order to get another syntax, called the \emph{decorated syntax}, 
such that the semantics \emph{is} a model of the decorated syntax. 
This approach leads to a new point of view about effects, 
which can be seen as a generalization of the point of view of \emph{monads}: 
the distinction between values and computations provided 
by the monad can be seen as a kind of decoration. 
In our framework every logic is sound (remark~\ref{remark:dialog-sound}),
and a computational effect is defined with respect to 
a \emph{span} of logics, which means, a pair of 
morphisms of logics with the same domain. 

\begin{defi}
\label{definition:dialog-effects}
Let $\zoom$ be a span in the category of simplified diagrammatic logics:  
  $$ \xymatrix@C=4pc@R=1pc{
  & \catT_\deco \ar[ld]_{F_\app} \ar[rd]^{F_\expl} & \\ 
  \catT_\app &  & \catT_\expl \\ 
  } $$
We call $\catT_\app$ the \emph{apparent} logic, 
$\catT_\deco$ the \emph{decorated} logic and 
$\catT_\expl$ the \emph{explicit} logic.
Let $G_\expl$ denote the right adjoint of $F_\expl$. 
A \emph{language with effect} with respect to $\zoom$ is a language 
$\Lang_\deco=(\Ff_\deco,\Tt_\deco,M_\deco)$ in $\catT_\deco$  
together with a theory $\Tt_\expl$ in $\catT_\expl$
such that $\Tt_\deco = G_\expl\Tt_\expl$. 
The \emph{apparent syntax} of $\Lang_\deco$ is 
$\Ff_\app=F_\app\Ff_\deco$ in $\catT_\app$.
The \emph{expansion} of $\Lang_\deco$ is the language 
$\Lang_\expl=(\Ff_\expl,\Tt_\expl,M_\expl)$ in $\catT_\expl$
with $\Ff_\expl=F_\expl\Ff_\deco$ and $M_\expl=\varphi M_\deco$, 
where $\varphi: \Mod_{\catT_\deco}(\Ff_\deco,\Tt_\deco) 
\to \Mod_{\catT_\expl}(\Ff_\expl,\Tt_\expl)$ 
is the bijection provided by the adjunction $F_\expl\dashv G_\expl$. 
\end{defi}
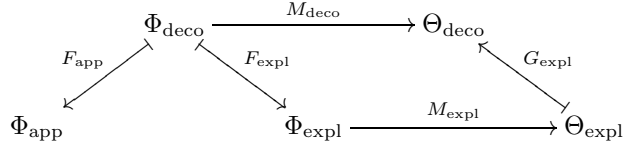
  $$ \xymatrix{
  & \Ff_\deco \ar@{|->}[dl]_{F_\app} \ar@{|->}[dr]^{F_\expl} \ar[rr]^{M_\deco} 
     && \Tt_\deco  & \\ 
  \Ff_\app && \Ff_\expl \ar[rr]^{M_\expl} && \Tt_\expl \ar@{|->}[ul]_{G_\expl}  \\ 
  }$$

\begin{rem}
\label{remark:dialog-effects-expl} 
Since a language with effect $\Lang_\deco$ is defined 
as a language on $\catT_\deco$, 
according to remark~\ref{remark:dialog-sound} it is \emph{sound}. 
Similarly, the expansion $\Lang_\expl$ of $\Lang_\deco$ is  
a language on $\catT_\expl$, hence it is \emph{sound}.  
Both languages are equivalent from the point of view of semantics,
thanks to the bijection $\varphi$.
This may be used for formalizing a computational effect 
when the decorated syntax corresponds to the programs  
while the explicit syntax does not, as in the bank account example
in section~\ref{subsec:bank}.  
\end{rem}

\begin{rem}
\label{remark:dialog-effects-app} 
It is tempting to look for a language $\Lang_\app=(\Ff_\app,\Tt_\app,M_\app)$ 
on $\catT_\app$, where $\Ff_\app=F_\app\Ff_\deco$
is the apparent syntax of $\Lang_\deco$. 
However, in general such a language does not exist 
(as for instance in remark~\ref{remark:states-effect}). 
\end{rem}

\section{States}
\label{sec:states}

In the syntax of an imperative language  
there is no type of states (the state is ``hidden'')  
while the interpretation of this language 
involves a set of states $\St$. 
More precisely, if the types $X$ and $Y$ 
are interpreted as the sets $\deno{X}$ and $\deno{Y}$, 
then each term $f:X\to Y$ is interpreted 
as a function $\deno{f}:\deno{X}\times \St \to \deno{Y}\times \St$. 
In Moggi's papers introducing monads for effects \cite{Mo89,Mo91}
such a term $f:X\to Y$ is called a \emph{computation},
and whenever the function $\deno{f}$ is $\deno{f}_\pure\times\id_{\St}$ 
for some $\deno{f}_\pure:\deno{X} \to \deno{Y}$ then $f$ is called 
a \emph{value}.
We keep this distinction, using \emph{modifier} and \emph{pure term}
instead of \emph{computation} and \emph{value}, respectively.
In addition, an \emph{accessor}
(or \emph{inspector}) is a term $f:X\to Y$ that is interpreted by a function 
$\deno{f}=\tuple{\deno{f}_\acc,\prr_{\deno{X}}}$,   
for some $\deno{f}_\acc:\deno{X}\times \St \to \deno{Y}$, 
where $\prr_{\deno{X}}:\deno{X}\times \St \to \St$ is the projection.   
It follows that every pure term is an accessor and every accessor is 
a modifier. 
We will use the decorations $\pure$, $\acc$ and $\modi$, 
written as superscripts, 
for pure terms, accessors and modifiers, respectively. 
Moreover, we distinguish two kinds of equations:
when $f,g:X\to Y$ are parallel terms, then 
a \emph{strong} equation $f \eqs g$ is interpreted as the equality 
$\deno{f}=\deno{g}: \deno{X}\times \St \to \deno{Y}\times \St$,
while a \emph{weak} equation $f \eqw g$ 
is interpreted as the equality 
$\prl_{\deno{Y}}\circ\deno{f}=\prl_{\deno{Y}}\circ\deno{g}: 
\deno{X}\times \St \to \deno{Y}$,
where $\prl_{\deno{Y}}:\deno{Y}\times \St \to \deno{Y}$ is the projection.  
Clearly, both notions coincide on accessors, hence on pure terms. 

\subsection{A span of logics for states}
\label{subsec:states-zoom}

Let $\Loc$ be a given set, called the set of \emph{locations}. 
Let us define a span of logics for dealing with states
(with respect to the set of locations $\Loc$) denoted by $\zoom_{\st}$: 
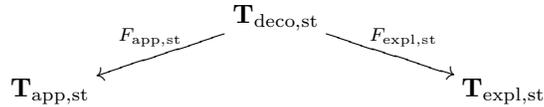
   $$ \xymatrix@C=4pc@R=1pc{
  & \catT_{\deco,\st} \ar[ld]_{F_{\app,\st}} \ar[rd]^{F_{\expl,\st}} & \\ 
  \catT_{\app,\st} &  & \catT_{\expl,\st} \\ 
  } $$
In this section the subscript ``$\st$'' will be omitted. 
First the decorated logic is defined, 
then the apparent logic and the morphism $F_\app$,
and finally the explicit logic and the morphism $F_\expl$. 
For each logic the definition of the morphisms of theories 
is omitted, since it derives in a natural way 
from the definition of the theories. 
In order to focus on the fundamental properties 
of states as effects, 
these logics are based on the 
\emph{monadic equational logic} (as in example~\ref{example:dialog-meqn}).

The logic $\catT_\deco$ is the 
\emph{decorated monadic equational logic for states}
(with respect to $\Loc$), defined as follows. 
A theory $\Tt_\deco$ for this logic is made of:
\begin{itemize}
\item Three nested monadic equational theories 
$\Tt^\pure \subseteq \Tt^\acc \subseteq \Tt^\modi$ with the same types,
such that the congruence on $\Tt^\pure$ and on $\Tt^\acc$
is the restriction of the congruence $\eqs$ on $\Tt^\modi$.
The objects of any of the three categories are called the \emph{types} of the theory,
the terms in $\Tt^\modi$ are called the \emph{modifiers}, 
those in $\Tt^\acc$ may be called the \emph{accessors}, 
and if they are in $\Tt^\pure$ they may be called the \emph{pure terms}. 
The relations $f \eqs g$ are called the \emph{strong equations}. 
\item An equivalence relation $\eqw$ between parallel terms, 
which satisfies the properties of substitution 
and pure replacement (defined in figure~\ref{fig:states-rules-deco}).
The relations $f \eqw g$ are called the \emph{weak equations}. 
Every strong equation is a weak equation and  
every weak equation between accessors is a strong equation. 
\item A distinguished type $\unit$  
which has the following \emph{decorated terminality} property: 
for each type $X$ there is a pure term $\tu_X:X\to \unit$ 
such that every modifier $f:X\to \unit$ satisfies $f \eqw \tu_X$. 
\item And $\Tt$ may have \emph{decorated products on $\Loc$}, 
where a decorated product on $\Loc$
is defined as a cone of accessors $(q_i:Y\to Y_i)_{i\in\Loc}$ 
such that for each cone of accessors $(f_i:X\to Y_i)_{i\in\Loc}$ 
with the same base
there is a modifier $\tuple{f_j}_{j\in\Loc}:X\to Y$ such that 
$q_i \circ \tuple{f_j}_{j\in\Loc} \eqw f_i$ for each $i$, and
whenever some modifier $g:X\to Y$ is such that 
$q_i \circ g \eqw f_i$ for each $i$  
then $g \eqs \tuple{f_j}_{j\in\Loc}$. 
\end{itemize}

Figure~\ref{fig:states-rules-deco} provides the \emph{decorated rules}
for states,   
which describe the properties of the decorated theories. 
We use the following conventions: 
$X,Y,Z,\dots$ are types, 
$f,g,h,\dots$ are terms, 
$f^\pure$ means that $f$ is a pure term, 
$f^\acc$ means that $f$ is an accessor,
and similarly $f^\modi$ means that $f$ is a modifier
(this is always the case but the decoration may be used for emphasizing).
Decoration hypotheses may be grouped with other hypotheses: 
for instance, ``$f^\acc \eqw g^\acc$''
means ``$f^\acc$ and $g^\acc$ and $f \eqw g$''. 
A decorated product on $\Loc$ is denoted by $(q_i^\acc:\prod_j Y_j\to Y_i)_i$. 

\begin{figure}[!h]
\renewcommand{\arraystretch}{3}
$$ \begin{array}{|c|} 
\hline 
\mbox{ Rules of the monadic equational logic, and: } \\ 
\rnpa \dfrac{f^\pure}{f^\acc} \qquad \rnam \dfrac{f^\acc}{f^\modi} \\ 
\rnpcomp \dfrac{f^\acc \quad g^\acc}{(g\circ f)^\acc} \qquad
\rnacomp \dfrac{f^\pure \quad g^\pure}{(g\circ f)^\pure}  \qquad 
\rnpid \dfrac{X}{\id_X^\pure:X\to X } \\ 
\rnwrefl \dfrac{}{f \eqw f} \qquad 
\rnwsym \dfrac{f \eqw g}{g \eqw f} \qquad 
\rnwtrans \dfrac{f \eqw g \quad g \eqw h}{f \eqw h} \\ 
\rnwsubs \dfrac{f:X\to Y \quad g_1\eqw g_2:Y\to Z}
  {g_1\circ f \eqw g_2\circ f :X\to Z}  \qquad 
\rnwrepl \dfrac{f_1\eqw f_2:X\to Y \quad g^\pure:Y\to Z}
  {g\circ f_1 \eqw g\circ f_2 :X\to Z} \\
\rnsw \dfrac{f \eqs g}{f \eqw g} \qquad
\rnws \dfrac{f^\acc \eqw g^\acc}{f \eqs g} \\
\rnfinal \dfrac{\quad}{\unit} \qquad  
\rnpfinal \dfrac{X}{\tu_X^\pure:X\to \unit} \qquad 
\rnwfinal \dfrac{f:X\to \unit}{f \eqw \tu_X} \\ 
\rntuple  \dfrac {(q_i^\acc\!:\!\prod_j\! Y_j\to Y_i)_i \quad 
    (f_i^\acc\!:\!X\to Y_i)_i} {\tuple{f_j}_j\!:\!X\to Y \quad
    \forall i\, q_i \circ \tuple{f_j}_j \eqw f_i } \\ 
\rntupleuns  \dfrac{(q_i^\acc\!:\!\prod_j\! Y_j\to Y_i)_i \quad 
    g:X\to Y \quad 
    \forall i\, q_i \circ g \eqw f_i^\acc}
  {g \eqs \tuple{f_j}_j} 
\rule[-20pt]{0pt}{0pt} \\
\hline 
\end{array}$$
\renewcommand{\arraystretch}{1}
\caption{Rules of the decorated logic for states}
\label{fig:states-rules-deco}
\end{figure}

\begin{rem}
There is no general replacement rule for weak equations:
if $f_1\eqw f_2:X\to Y$ and $g:Y\to Z$ 
then in general $g\circ f_1 \not\eqw g\circ f_2$,
except when $g$ is pure. 
\end{rem}

\begin{exa}
\label{example:toterminal} 
Let us derive the following rule, 
which says that $\tu_X$ is the unique accessor from $X$ to $\unit$,  
up to strong equations: 
 $$ \rnsfinal \frac{f^\acc:X\to \unit }{f \eqs \tu_X} $$
The derivation tree is: 
\begin{prooftree}
\AxiomC{$f^\acc$}
    \AxiomC{$f:X\to \unit$}
    \LeftLabel{\rnwfinal} 
    \UnaryInfC{$f \eqw \tu_X$}
          \AxiomC{$X$}
          \LeftLabel{\rnpfinal} 
          \UnaryInfC{$\tu_X^\pure$}
          \LeftLabel{\rnpa} 
          \UnaryInfC{$\tu_X^\acc$}
      \LeftLabel{\rnws} 
      \TrinaryInfC{$f \eqs \tu_X$}
\end{prooftree}
\end{exa}

Now let us describe the ``apparent'' side of the span. 
The logic $\catT_\app$ extends the monadic equational logic as follows~:
a theory of $\catT_\app$ is a monadic equational theory 
with a terminal object $\unit$ 
which may have products on $\Loc$ (i.e., with their base indexed by $\Loc$). 
The morphism $F_\app:\catT_\deco\to\catT_\app$ maps
each theory $\Tt_\deco$ of $\catT_\deco$ 
to the theory $\Tt_\app$ of $\catT_\app$ made of: 
  \begin{itemize}  
  \item A type $\ha{X}$ for each type $X$ in $\Tt_\deco$. 
  \item A term $\ha{f}:\ha{X}\to \ha{Y}$ for each modifier $f:X\to Y$ 
    in $\Tt_\deco$
  (which includes the accessors and the pure terms), 
  such that   
  $\ha{\id_X}=\id_{\ha{X}}$ for each type $X$ and 
  $\ha{g\circ f}=\ha{g}\circ\ha{f}$ for each pair of 
  consecutive modifiers $(f,g)$.
  \item An equation $\ha{f}\eqs \ha{g}$ 
    for each weak equation $f\eqw g$ in $\Tt_\deco$
  (which includes the strong equations). 
  \item A product $(\ha{q_i}:\prod_j\ha{Y_j}\to \ha{Y_i})_{i\in\Loc}$
  for each decorated product $(q_i^\acc:\prod_j Y_j\to Y_i)_{i\in\Loc}$ 
    in $\Tt_\deco$. 
  \end{itemize}
Thus, the morphism $F_\app$ blurs the distinction between 
modifiers, accessors and pure terms,
as well as the distinction between weak and strong equations. 
In the following, the notation $\ha{\dots}$ will be omitted. 

It follows from the definition of $F_\app$ that 
each rule of the decorated logic $\catT_\deco$ is mapped 
by $F_\app$ to a rule of the apparent logic $\catT_\app$,
so that $F_\app$ is a morphism of diagrammatic logics. 
The morphism $F_\app$ can be used for checking 
a decorated proof in two steps, 
by checking first that its
image by $F_\app$ is a proof in $\catT_\app$. 

Now let us describe the ``explicit'' side of the span. 
The logic $\catT_\expl$ extends the monadic equational logic as follows~: 
a theory of $\catT_\expl$ is a monadic equational theory with a 
distinguished object $S$, called the \emph{type of states},
with a product-with-$S$ functor $X\times S$,  
and which may have products on $\Loc$. 
The morphism $F_\expl:\catT_\deco \to \catT_\expl$ maps  
each theory $\Tt_\deco=(\Tt^\pure\subseteq \Tt^\acc\subseteq \Tt^\modi)$ 
of $\catT_\deco$ 
to the theory $\Tt_\expl$ of $\catT_\expl$ made of: 
  \begin{itemize}
  \item A type $\ti{X}$ for each type $X$ in $\Tt_\deco$;
  the projections from $\ti{X}\times S$ are denoted by
  $\prl_X:\ti{X}\times S \to \ti{X}$ and $\prr_X:\ti{X}\times S \to S$.
  \item A term $\ti{f}:\ti{X}\times S\to \ti{Y}\times S$ for each modifier 
  $f:X\to Y$ in $\Tt_\deco$, such that: 
    \begin{itemize}
    \item if in addition $f$ is an accessor 
    then there is a term $\ti{f}_\acc:\ti{X}\times S\to \ti{Y}$
    such that $\ti{f}=\tuple{\ti{f}_\acc,\prr_X}$,
    \item and if moreover $f$ is a pure term  
    then there is a term $\ti{f}_\pure:\ti{X}\to \ti{Y}$
    such that $\ti{f}_\acc=\ti{f}_\pure\circ\prl_X: \ti{X}\times S\to \ti{Y}$,
    hence $\ti{f}=\tuple{\ti{f}_\pure\circ\prl_X,\prr_X}=\ti{f}_\pure\times\id_S$.
    \end{itemize}
    such that 
    $\ti{\id_X}=\id_{\ti{X}\times S}$ for each type $X$ and 
    $\ti{g\circ f}=\ti{g}\circ\ti{f}$ for each pair of 
    consecutive modifiers $(f,g)$. 
  \item An equation $\ti{f} \eqs \ti{g} : \ti{X}\times S\to \ti{Y}\times S$
  for each strong equation $f \eqs g : X\to Y$ in $\Tt_\deco$. 
  \item An equation $\prl_Y \circ \ti{f} \eqs \prl_Y \circ \ti{g} : \ti{X}\times S\to \ti{Y}$
  for each weak equation $f \eqw g : X\to Y$ in $\Tt_\deco$. 
  \item A product $((\ti{q_i})_\acc:(\prod_j Y_j)\times S\to Y_i)_{i\in\Loc}$ 
  for each decorated product $(q_i^\acc:\prod_j Y_j\to Y_i)_{i\in\Loc}$ in $\Tt_\deco$.  
  \end{itemize}
Thus, the morphism $F_\expl$ makes explicit the meaning of the decorations,
by introducing a ``type of states'' $S$.
In the following, the notation $\ti{\dots}$ will sometimes be omitted
(mainly for types). 
The morphism $F_\expl$ is such that 
each modifier $f$ gives rise to a term $\ti{f}$ 
which may use and modify the state,
while whenever $f$ is an accessor then $\ti{f}$ may use the state 
but is not allowed to modify it,
and when moreover $f$ is a pure term then $\ti{f}$ 
may neither use nor modify the state.
When $f \eqs g$ then $\ti{f}$ and $\ti{g}$ 
must return the same result and the same state;  
when $f \eqw g$ then $\ti{f}$ and $\ti{g}$ 
must return the same result but maybe not the same state. 
 
\begin{rem}
\label{remark:states-comp}
When $f$ and $g$ are consecutive modifiers, we have defined 
$\ti{g\circ f}=\ti{g}\circ \ti{f}$. 
Thus, when $f$ and $g$ are accessors, 
the accessor $g\circ f$ is such that 
$\ti{g\circ f}=\tuple{\ti{g}_\acc,\prr_Y}\circ \ti{f}
=\tuple{\ti{g}_\acc\circ \ti{f},\prr_Y\circ \ti{f}}
=\tuple{\ti{g}_\acc\circ\ti{f},\prr_X}$,
so that $\ti{g\circ f}_\acc = \ti{g}_\acc\circ\ti{f}$: 
we recognize the co-Kleisli composition 
of $\ti{f}_\acc$ and $\ti{g}_\acc$ with respect to the comonad $-\times S$.
When $f$ and $g$ are pure then the pure term $g\circ f$ 
is such that $\ti{g\circ f}_\pure =\ti{g}_\pure \circ \ti{f}_\pure$.
\end{rem}

Altogether, the span of logics for states $\zoom_{\st}$ 
is summarized in figure~\ref{fig:states-zoom}.

\begin{figure}[!h]
\renewcommand{\arraystretch}{1.1}
$$ \begin{array}{|cc|c|cc|}
\hline 
\multicolumn{1}{|c}{\catT_\app} & 
\multicolumn{1}{c}{\lupto{F_\app}} & 
\multicolumn{1}{c}{\catT_\deco } &
\multicolumn{1}{c}{\rupto{F_\expl}} &  
\multicolumn{1}{c|}{\catT_\expl} \\
\hline 
&& \mbox{ modifier } && \\
f:X\to Y && 
f:X\to Y  && 
\ti{f}:X\times S\to Y\times S  \\
&& \mbox{ accessor } && \\
f:X\to Y && 
f^\acc:X\to Y  && 
\ti{f}_\acc:X\times S\to Y  \\
&& \mbox{ pure term } && \\
f:X\to Y && 
f^\pure:X\to Y  && 
\ti{f}_\pure:X\to Y  \\
\hline 
&& \mbox{ strong equation } && \\ 
f \eqs g:X\to Y && 
f \eqs g:X\to Y && 
\ti{f} \eqs \ti{g} : X\times S\to Y\times S  \\
&& \mbox{ weak equation } && \\
f \eqs g:X\to Y && 
f \eqw g:X\to Y && 
\prl_Y\circ \ti{f} \eqs \prl_Y\circ \ti{g} : X\times S\to Y  \\
\hline 
\end{array} $$
\renewcommand{\arraystretch}{1}
\caption{The span of logics for states}
\label{fig:states-zoom} 
\end{figure}

\subsection{States as effect}
\label{subsec:states-effect}

Now let us introduce the operations and equations 
related to the states effect.  
We consider the semantics of states as the semantics 
of a language with effect,
in the sense of definition~\ref{definition:dialog-effects}, 
with respect to the span of logics for states $\zoom_{\st}$
defined in section~\ref{subsec:states-zoom}. 
This language with effect 
$\Lang_{\deco,\st}=(\Ff_{\deco,\st},\Tt_{\deco,\st},M_{\deco,\st})$
is defined below (the index ``$\st$'' is omitted) in the following way: 
\begin{itemize}
\item first the apparent syntax $\Ff_\app$,
the decorated syntax $\Ff_\deco$ 
and the explicit syntax $\Ff_\expl=F_\expl\Ff_\deco$; 
\item then the explicit theory $\Tt_\expl$ and 
the explicit semantics $M_\expl:\Ff_\expl \to \Tt_\expl$,
which form the expansion $\Lang_\expl$ of $\Lang_\deco$;
\item and finally the decorated theory $\Tt_\deco=G_\expl\Tt_\expl$
and the decorated semantics $M_\deco = \varphi^{-1}M_\expl$, 
where $\varphi: \Mod_{\catT}(\Ff,\Tt) \to \Mod_{\catT'}(\Ff',\Tt') $
is the bijection provided by the adjunction $F\dashv G$. 
\end{itemize}

The apparent syntax $\Ff_\app$ is built as follows.
For each location $i$ there is 
a type $V_i$ for the possible values of $i$ 
and an operation $\fl_i:\unit \to V_i$ for observing the value of $i$. 
These operations form a product on $\Loc$
$(\fl_i:\unit\to V_i)_{i\in\Loc}$, 
so that for each location $i$ there is an operation 
$\fu_i:V_i\to \unit$, unique up to congruence, 
which satisfies the equations
  $$ \begin{cases}
  \fl_i\circ \fu_i \eqs \id_{V_i} & : V_i\to V_i \\ 
  \fl_j\circ \fu_i \eqs \fl_j\circ \tu_{V_i} & : V_i\to V_j \; \mbox{ for each }j\ne i \\
  \end{cases}$$
Intuitively, this means that after $\fu_i(a)$ is executed, 
the value of $i$ is put to $a$ and the value of $j$ (for $j\ne i$) is unchanged. 

\begin{rem}
\label{remark:states-effect} 
Let $\Tt_\app$ be the category of sets seen as a theory 
of the apparent logic (with equality as congruence). 
Let us try to build progressively a model of $\Ff_\app$ in $\Tt_\app$. 
The type $\unit$ must be interpreted as a singleton $\{*\}$,
and for each $i$ the interpretation of $V_i$ is a set $\Val_i$.
Thus, the interpretation of $\fl_i$ is an element of $\Val_i$,
and each interpretation of the $V_i$'s and $\fl_i$'s in $\Tt_\app$ 
corresponds to a state, made of a value for each location; 
this is known as the \emph{states-as-models} or \emph{states-as-algebras}  
point of view \cite{GDK96}. 
This interpretation can be extended to $\fu_i:V_i\to\unit$ in only one way: 
indeed $\fu_i$ must be interpreted as the function which maps 
every value in $\Val_i$ to $*$. It follows that,
as soon as the set $\Val_i$ is not a singleton, 
the equation $\fl_i\circ \fu_i \eqs \id_{V_i}$ cannot be satisfied. 
Thus, the intended semantics of states 
cannot be a model of the apparent syntax $\Ff_\app$ in $\Tt_\app$,
as mentioned in remark~\ref{remark:dialog-effects-app}.
\end{rem}

The decorated syntax $\Ff_\deco$ is obtained by adding informations 
(decorations) to $\Ff_\app$. It is generated by 
a type $V_i$  
and an accessor $\fl_i^\acc:\unit\to V_i$ for each $i\in\Loc$, 
which form a decorated product $(\fl_i^\acc:\unit\to V_i)_{i\in\Loc}$. 
The operations $\fu_i$'s are decorated as modifiers
and the equations as weak equations: 
  \begin{equation}
  \label{eq:states} \begin{cases}
  \fl_i^\acc \circ \fu_i^\modi \eqw \id_{V_i}^\pure & : V_i\to V_i \\ 
  \fl_j^\acc \circ \fu_i^\modi \eqw \fl_j^\acc \circ \tu_{V_i}^\pure & : V_i\to V_j 
    \; \mbox{ for each }j\ne i \\
  \end{cases}
  \end{equation}
It follows from the rules of the decorated logic 
that in every decorated theory there is an interpretation 
for the $\fu_i$'s, which is unique up to strong equations.
As required, the apparent syntax $\Ff_\app=F_\app\Ff_\deco$
is recovered by dropping the decorations. 

Using the definition of $F_\expl$ in section~\ref{subsec:states-zoom}, 
we get the explicit syntax $\Ff_\expl=F_\expl\Ff_\deco$.
It is the theory in the explicit logic generated by 
a type $V_i$  
and a term $\ti{\fl_i}_\acc : S\to V_i$ for each $i\in\Loc$, 
which form a product $(\ti{\fl_i}_\acc :S\to V_i)_{i\in\Loc}$.
So, for each location $i$, the operation $\ti{\fu_i}:V_i\times S\to S$
is defined up to congruence by the equations: 
  $$ \begin{cases}
  \ti{\fl_i}_\acc\circ \ti{\fu_i} \eqs \prl_{V_i} & : V_i\times S\to V_i \\ 
  \ti{\fl_j}_\acc\circ \ti{\fu_i} \eqs \ti{\fl_j}_\acc\circ \prr_{V_i} & 
    : V_i\times S\to V_j \; \mbox{ for each }j\ne i \\
   \end{cases}$$
   
The explicit theory $\Tt_\expl$ is made of the category of sets 
with the equality as congruence, 
with a distinguished set $\St$ called the set of states, 
with cartesian products with $\St$, 
and with a product on $\Loc$
with vertex $\St$, denoted by $(\sfl_i:\St\to \Val_i)_{i\in\Loc}$,  
so that $\St=\prod_{j\in\Loc} \Val_j$.  
The explicit semantics $M_\expl:\Ff_\expl\to\Tt_\expl$ is the model 
(in the explicit logic) which maps 
$S$ to $\St$ and, for each $i\in\Loc$,  
the type $V_i$ to the set $\Val_i$
and the operations $\fl_i$ and $\fu_i$
to the functions $\sfl_i$ and $\sfu_i$, respectively. 

The decorated semantics $M_\deco:\Ff_\deco\to\Tt_\deco$ 
is obtained from the explicit semantics $M_\expl:\Ff_\expl\to\Tt_\expl$
thanks to the adjunction $F_\expl\dashv G_\expl$. 
The decorated theory $\Tt_\deco=G_\expl\Tt_\expl$ 
has a type for each set, 
a modifier $f^\modi:X\to Y$ for each function $f:X\times \St \to Y\times \St$,
an accessor $f^\acc:X\to Y$ for each function $f:X\times \St \to Y$ 
and a pure term $f^\pure:X\to Y$ for each function $f:X \to Y$,
with the straightforward conversions. 
It follows that there are in $\Tt_\deco$, for each $i\in\Loc$,
an accessor $\fl_i^\acc:\unit\to\Val_i$ and 
a modifier $\fu_i^\modi:\Val_i\to\unit$, 
and that we get the model $M_\deco=\varphi^{-1}M_\expl$ by mapping 
the type $V_i$ to the set $\Val_i$, 
the accessor $\fl_i^\acc$ to the function $\sfl_i$ 
and the modifier $\fu_i^\modi$ to the function $\sfu_i$, 
for each $i\in\Loc$. 

According to claim~\ref{claim:states-explicit}, 
the explicit model $M_\expl$ provides the intended semantics of states. 
By adjunction, this is also the semantics of the decorated model $M_\deco$,
hence the following result.

\begin{prop}
\label{proposition:states}
The language with effect $\Lang_{\deco,\st}$ 
provides the intended semantics of states. 
\end{prop}

To conclude this section,
the decorated logic is used for proving a 
fundamental property of states: 
when a state $s$ is modified by updating a location $i$
with its own value in $s$, 
then the resulting state is undistinguishable from $s$;
this is the first of equations~\ref{eq:states-seven}.
It should be reminded that each decorated proof may be mapped to 
an equational proof either by dropping the decorations 
(using the morphism $F_\app$)
or by expliciting them (using the morphism $F_\expl$). 
In the first case one gets a correct proof which may be quite uninteresting, 
in the second case one gets a correct proof which may be quite complicated. 

\begin{prop}
\label{proposition:states-prop}
For every $i\in\Loc$:
$$ \begin{cases}
 \fu_i^\modi \circ \fl_i^\acc \eqs \id_\unit^\pure & 
  \mbox{ in the decorated logic } \\ 
 \ti{\fu_i} \circ \ti{\fl_i} \eqs \id_S  & 
  \mbox{ in the explicit logic } \\ 
\end{cases} $$
\end{prop}

\proof 
In the decorated logic, let us prove the weak equations 
$ \fl_j \circ \fu_i \circ \fl_i \eqw \fl_j $ for each $j\in\Loc$; 
then the first result will follow from the rule for decorated products on $\Loc$
and the second result by applying the morphism $F_\expl$.
In the following decorated proofs, the rules for associativity 
and identities are omitted.
When $j=i$, the substitution rule for $\eqw$ yields: 
\footnotesize
\begin{prooftree}  
\AxiomC{$\fl_i \circ \fu_i  \eqw \id_{V_i} $} 
\LeftLabel{\rnwsubs}
\UnaryInfC{$\fl_i \circ \fu_i \circ \fl_i \eqw \fl_i$}
\end{prooftree}
\normalsize
When $j\ne i$, using the substitution rule for $\eqw$  
and the replacement rule for $\eqs$ we get: 
\footnotesize
\begin{prooftree}  
\AxiomC{$\fl_j \circ \fu_i  \eqw \fl_j \circ \tu_{V_i} $} 
\LeftLabel{\rnwsubs}
\UnaryInfC{$\fl_j \circ \fu_i \circ \fl_i \eqw \fl_j \circ \tu_{V_i} \circ \fl_i $}
        \AxiomC{\vdots}
        \UnaryInfC{$\tu_{V_i} \circ \fl_i \eqs \id_{\unit} $}  
        \LeftLabel{\rnsrepl} 
        \UnaryInfC{$\fl_j \circ \tu_{V_i} \circ \fl_i \eqs \fl_j $} 
        \LeftLabel{\rnsw}
        \UnaryInfC{$\fl_j \circ \tu_{V_i} \circ \fl_i \eqw \fl_j $}
    \LeftLabel{\rnwtrans}
    \BinaryInfC{$\fl_j \circ \fu_i \circ \fl_i \eqw \fl_j$}
\end{prooftree}
\normalsize
\qed

\section{Exceptions}
\label{sec:exceptions}

It has been seen in section~\ref{sec:explicit} that there is 
a duality between the semantics of states and 
the semantics of exceptions. 
A decorated language for states as effects 
has been designed in section~\ref{sec:states}. 
Now, in section~\ref{subsec:exceptions-dual}, 
we define a decorated language for exceptions as effects 
simply by dualizing section~\ref{sec:states}; 
this provides the key operations for exceptions. 
Then in section~\ref{subsec:exceptions-encaps} 
we check that the encapsulation of the key functions 
from section~\ref{sec:explicit} may be performed in the 
decorated syntax. 

\subsection{Dualizing states}
\label{subsec:exceptions-dual}

Let us dualize section~\ref{sec:states}.
Let $\Const$ be a given set, called the set of \emph{exception constructors}. 
The span of logics for dealing with exceptions
(with respect to $\Const$) is denoted by $\zoom_{\exc}$: 
   $$ \xymatrix@C=4pc@R=1pc{
  & \catT_{\deco,\exc} \ar[ld]_{F_{\app,\exc}} \ar[rd]^{F_{\expl,\exc}} & \\ 
  \catT_{\app,\exc} &  & \catT_{\expl,\exc} \\ 
  } $$
In this section the subscript ``$\exc$'' will be omitted. 
In order to focus on the fundamental properties of exceptions as effects, 
these logics are based on the \emph{monadic equational logic}.
Some additional features will be added in section~\ref{subsec:exceptions-encaps}.

A theory for the 
\emph{decorated monadic equational logic for exceptions} $\catT_\deco$
is made of:
\begin{itemize}
\item Three nested monadic equational theories 
$\Tt^\pure \subseteq \Tt^\acc \subseteq \Tt^\modi$ with the same types,
such that the congruence on $\Tt^\pure$ and on $\Tt^\acc$
is the restriction of the congruence $\eqs$ on $\Tt^\modi$.
The objects of any of the three categories are called the \emph{types} of the theory,
the terms in $\Tt^\modi$ are the \emph{catchers}, 
those in $\Tt^\acc$ are the \emph{propagators} (or \emph{throwers}) 
and those in $\Tt^\pure$ are the \emph{pure terms}. 
The relations $f \eqs g$ are called the \emph{strong equations}. 
\item An equivalence relation $\eqw$ between parallel terms, 
which satisfies the properties of replacement 
and pure substitution (as in figure~\ref{fig:exceptions-rules-deco}).
The relations $f \eqw g$ are called the \emph{weak equations}. 
Every strong equation is a weak equation and  
every weak equation between propagators is a strong equation. 
\item A distinguished type $\empt$  
which has the following \emph{decorated initiality} property: 
for each type $X$ there is a pure term $\cotu_X:\empt\to X$ 
such that every catcher $f:\empt\to X$ satisfies $f \eqw \cotu_X$. 
\item And $\Tt$ may have \emph{decorated coproducts on $\Const$},
i.e., cocones of propagators $(q_i:X_i\to X)_{i\in\Const}$ 
such that for each cocone of propagators $(f_i:X_i\to Y)_{i\in\Const}$ 
with the same base
there is a catcher $\cotuple{f_j}_{j\in\Const}:X\to Y$ such that 
$\tuple{f_i}_{i\in\Const} \circ q_i \eqw f_i$ for each $i$, and
whenever some catcher $g:X\to Y$ is such that 
$g\circ q_i \eqw f_i$ for each $i$  
then $g \eqs \cotuple{f_j}_{j\in\Const}$. 
\end{itemize}

Figure~\ref{fig:exceptions-rules-deco} provides the \emph{decorated rules}
for exceptions,   
which describe the properties of the decorated theories. 
We use the following conventions: 
$X,Y,Z,\dots$ are types, 
$f,g,h,\dots$ are terms, 
$f^\pure$ means that $f$ is a pure term, 
$f^\ppg$ means that $f$ is a propagator,
and similarly $f^\ctc$ means that $f$ is a catcher
(used for emphasizing). 
A decorated coproduct on $\Const$ is denoted by 
$(q_i^\ppg:X_i \to \sum_j X_j)_i$. 

\begin{figure}[!h]
\renewcommand{\arraystretch}{3}
$$ \begin{array}{|c|} 
\hline 
\mbox{ Rules of the monadic equational logic, and: } \\ 
\dfrac{f^\pure}{f^\acc} \qquad \dfrac{f^\acc}{f^\modi} \\ 
\dfrac{f^\acc \quad g^\acc}{(g\circ f)^\acc} \qquad
\dfrac{f^\pure \quad g^\pure}{(g\circ f)^\pure}  \qquad 
\dfrac{X}{\id_X^\pure:X\to X } \\ 
\dfrac{}{f \eqw f} \qquad 
\dfrac{f \eqw g}{g \eqw f} \qquad 
\dfrac{f \eqw g \quad g \eqw h}{f \eqw h} \\ 
\dfrac{f^\pure:X\to Y \quad g_1\eqw g_2:Y\to Z}
  {g_1\circ f \eqw g_2\circ f :X\to Z}  \qquad 
\dfrac{f_1\eqw f_2:X\to Y \quad g:Y\to Z}
  {g\circ f_1 \eqw g\circ f_2 :X\to Z} \\
\dfrac{f \eqs g}{f \eqw g} \qquad
\dfrac{f^\ppg \eqw g^\ppg}{f \eqs g} \\
\dfrac{\quad}{\empt} \qquad  
\dfrac{X}{\cotu_X^\pure:\empt\to X} \qquad 
\dfrac{f:\empt\to X}{f \eqw \cotu_X} \\ 
  \dfrac {(q_i^\ppg\!:\!X_i \to \sum_j\! X_j)_i \;\; 
    (f_i^\ppg\!:\!X_i\to Y)_i} {\cotuple{f_j}_j\!:\!X\to Y \;\;
    \forall i\, \tuple{f_i}_i \circ q_i \eqw f_i } \qquad 
  \dfrac{(q_i^\acc\!:\!X_i \to \sum_j\! X_j)_i \;\;
    g:X\to Y \;\;
    \forall i\, g\circ q_i \eqw f_i^\ppg}
  {g \eqs \cotuple{f_j}_j} 
\rule[-20pt]{0pt}{0pt} \\
\hline 
\end{array}$$
\renewcommand{\arraystretch}{1}
\caption{Rules of the decorated logic for exceptions}
\label{fig:exceptions-rules-deco}
\end{figure}

\begin{rem}
There is no general substitution rule for weak equations:
if $f:X\to Y$ and $g_1\eqw g_2:Y\to Z$ 
then in general $g_1\circ f \not\eqw g_2\circ f$,
except when $f$ is pure. 
\end{rem}

On the ``apparent'' side of the span, 
a theory for the apparent logic $\catT_\app$ is a monadic equational theory 
with an initial object $\empt$ 
which may have coproducts on $\Const$.
The morphism $F_\app:\catT_\deco\to\catT_\app$ maps
each theory $\Tt_\deco$ of $\catT_\deco$ 
to the theory $\Tt_\app$ of $\catT_\app$ made of: 
  \begin{itemize}  
  \item A type $\ha{X}$ for each type $X$ in $\Tt_\deco$. 
  \item A term $\ha{f}:\ha{X}\to \ha{Y}$ for each catcher $f:X\to Y$ in $\Tt_\deco$
  (which includes the propagators and the pure terms), 
  such that   
  $\ha{\id_X}=\id_{\ha{X}}$ for each type $X$ and 
  $\ha{g\circ f}=\ha{g}\circ\ha{f}$ for each pair of 
  consecutive catchers $(f,g)$.
  \item An equation $\ha{f}\eqs \ha{g}$ for each weak equation $f\eqw g$ in $\Tt_\deco$
  (which includes the strong equations). 
  \item A coproduct $(\ha{q_i}:\ha{X_i} \to \sum_j\ha{X_j})_{i\in\Const}$
  for each decorated coproduct $(q_i^\ppg:X_i \to \prod_j X_j)_{i\in\Const}$ in $\Tt_\deco$. 
  \end{itemize}
Thus, the morphism $F_\app$ blurs the distinction between 
catchers, propagators and pure terms,
and the distinction between weak and strong equations. 
In the following, the notation $\ha{\dots}$ will be omitted. 

On the ``explicit'' side of the span,
a theory for the explicit logic $\catT_\expl$ is a monadic equational theory 
with a distinguished object $E$, called the \emph{type of exceptions},
with a coproduct-with-$E$ functor $X+E$,  
and which may have coproducts on $\Const$. 
The morphism $F_\expl:\catT_\deco \to \catT_\expl$ maps  
each theory $\Tt_\deco=(\Tt^\pure\subseteq \Tt^\ppg\subseteq \Tt^\ctc)$ 
of $\catT_\deco$ 
to the theory $\Tt_\expl$ of $\catT_\expl$ made of: 
  \begin{itemize}
  \item A type $\ti{X}$ for each type $X$ in $\Tt_\deco$;
  the coprojections in $\ti{X}+E$ are denoted by 
  $\inl_X:\ti{X} \to \ti{X}+E$ and $\inr_X:E \to \ti{X}+E$.
  \item A term $\ti{f}:\ti{X}+E\to \ti{Y}+E$ for each catcher 
  $f:X\to Y$ in $\Tt_\deco$, such that: 
    \begin{itemize}
    \item if in addition $f$ is a propagator  
    then there is a term $\ti{f}_\ppg:\ti{X}\to \ti{Y}+E$
    such that $\ti{f}=\cotuple{\ti{f}_\ppg|\inr_Y}$,
    \item and if moreover $f$ is a pure term 
    then there is a term $\ti{f}_\pure:\ti{X}\to \ti{Y}$
    such that $\ti{f}_\ppg=\inl_Y\circ \ti{f}_\pure: \ti{X}\to \ti{Y}+E$,
    hence $\ti{f}=\cotuple{\inl_Y\circ\ti{f}_\pure|\inr_X}=\ti{f}_\pure+\id_E$.
    \end{itemize}
  and such that 
  $\ti{\id_X}=\id_{\ti{X}+E}$ for each type $X$ and 
  $\ti{g\circ f}=\ti{g}\circ\ti{f}$ for each pair of 
  consecutive catchers $(f,g)$.
  \item An equation $\ti{f} \eqs \ti{g} : \ti{X}+E\to \ti{Y}+E$
  for each strong equation $f \eqs g : X\to Y$ in $\Tt_\deco$. 
  \item An equation $\ti{f} \circ \inl_X \eqs \ti{g} \circ \inl_Y : 
    \ti{X}\to \ti{Y}+E$
  for each weak equation $f \eqw g : X\to Y$ in $\Tt_\deco$. 
  \item A coproduct $((\ti{q_i})_\ppg:(X_i \to (\sum_j X_j)+E)_{i\in\Const}$ 
  for each decorated coproduct $(q_i^\ppg:X_i\to \sum_j X_j)_{i\in\Const}$ 
    in $\Tt_\deco$.  
  \end{itemize}
Thus, the morphism $F_\expl$ makes explicit the meaning of the decorations,
by introducing a ``type of exceptions'' $E$ which does not appear in the syntax. 
In the following, the notation $\ti{\dots}$ will sometimes be omitted
(mainly for types). 
The morphism $F_\expl$ is such that 
each catcher $f$ gives rise to a term $\ti{f}$ 
which does not distinguish exceptions from ordinary values,  
while whenever $f$ is a propagator then $\ti{f}$ may 
throw an exception but it must propagate exceptions,  
and when moreover $f$ is a pure term then $\ti{f}$ 
must turn an ordinary value to an ordinary value and 
it must propagate exceptions.  
When $f \eqs g$ then $\ti{f}$ and $\ti{g}$ 
must coincide on ordinary values and on exceptions;  
when $f \eqw g$ then $\ti{f}$ and $\ti{g}$ 
must coincide on ordinary values but maybe not on exceptions. 
 
\begin{rem}
\label{remark:exceptions-comp}
When $f$ and $g$ are consecutive catchers, we have defined 
$\ti{g\circ f}=\ti{g}\circ \ti{f}$. 
Thus, dually to remark~\ref{remark:states-comp}, 
when $f$ and $g$ are propagators then the propagator $g\circ f$ 
is such that $\ti{g\circ f}_\ppg$ is the Kleisli composition 
of $\ti{f}_\ppg$ and $\ti{g}_\ppg$ with respect to the monad $-+E$, 
and when $f$ and $g$ are pure then the pure term $g\circ f$ 
is such that $\ti{g\circ f}_\pure =\ti{g}_\pure \circ \ti{f}_\pure$.
\end{rem}

Altogether, the span of logics for exceptions $\zoom_{\exc}$ 
is summarized in figure~\ref{fig:exceptions-zoom}.

\begin{figure}[!h]
\renewcommand{\arraystretch}{1.1}
$$ \begin{array}{|cc|c|cc|}
\hline 
\multicolumn{1}{|c}{\catT_\app} & 
\multicolumn{1}{c}{\lupto{F_\app}} & 
\multicolumn{1}{c}{\catT_\deco } &
\multicolumn{1}{c}{\rupto{F_\expl}} &  
\multicolumn{1}{c|}{\catT_\expl} \\
\hline 
&& \mbox{ catcher } && \\
f:X\to Y && 
f:X\to Y  && 
\ti{f}:X+E \to Y+E  \\
&& \mbox{ propagator } && \\
f:X\to Y && 
f^\ppg:X\to Y  && 
\ti{f}_\ppg:X\to Y+E  \\
&& \mbox{ pure term } && \\
f:X\to Y && 
f^\pure:X\to Y  && 
\ti{f}_\pure:X\to Y  \\
\hline 
&& \mbox{ strong equation } && \\ 
f \eqs g:X\to Y && 
f \eqs g:X\to Y && 
\ti{f} \eqs \ti{g} : X+E\to Y+E  \\
&& \mbox{ weak equation } && \\
f \eqs g:X\to Y && 
f \eqw g:X\to Y && 
\ti{f}\circ\inl_X \eqs \ti{g}\circ\inl_X : X\to Y+E  \\
\hline 
\end{array} $$
\renewcommand{\arraystretch}{1}
\caption{The span of logics for exceptions}
\label{fig:exceptions-zoom} 
\end{figure}

Now we consider the semantics of exceptions as the semantics 
of a language with effect 
$\Lang_{\deco,\exc}=(\Ff_{\deco,\exc},\Tt_{\deco,\exc},M_{\deco,\exc})$
with respect to the span of logics $\zoom_{\exc}$.

The apparent syntax $\Ff_\app$ is built as follows.
For each exception constructor $i$ there is 
a type $P_i$ for the possible parameters 
and an operation $\ft_i:P_i\to\empt$ called the \emph{key thrower},
for throwing an exception of constructor $i$. 
These operations form a coproduct on $\Const$
$(\ft_i:P_i\to\empt)_{i\in\Const}$, 
so that for each $i$ there is an operation 
$\fc_i:\empt\to P_i$, unique up to congruence), 
called the \emph{key catcher}, 
which satisfies the equations
  $$ \begin{cases}
  \fc_i\circ \ft_i \eqs \id_{P_i} & : P_i\to P_i \\ 
  \fc_i\circ \ft_j \eqs \cotu_{P_i}\circ \ft_j  & : P_j\to P_i \; \mbox{ for each }j\ne i \\
  \end{cases}$$
Intuitively, this means that when $\fc_i$ is called, 
the parameter of the previous call to $\ft_i$ (for the same $i$) is returned. 

The decorated syntax $\Ff_\deco$ is obtained by adding informations 
(decorations) to $\Ff_\app$. It is generated by 
a type $P_i$  
and a propagator $\ft_i^\ppg:P_i\to\empt$ for each $i\in\Const$, 
which form a decorated coproduct $(\ft_i^\ppg:P_i\to\empt)_{i\in\Const}$. 
The operations $\fc_i$'s are decorated as catchers
and the equations as weak equations: 
  \begin{equation}
  \label{eq:exceptions} 
  \begin{cases}
  \fc_i^\ctc \circ \ft_i^\ppg \eqw \id_{P_i}^\pure & : P_i\to P_i \\ 
  \fc_i^\ctc \circ \ft_j^\ppg \eqw \cotu_{P_i}^\pure \circ \ft_j^\ppg & : P_j\to P_i 
    \; \mbox{ for each }j\ne i \\
  \end{cases}
  \end{equation}
It follows from the rules of the decorated logic 
that in every decorated theory there is an interpretation 
for the $\fc_i$'s, which is unique up to strong equations. 
The apparent syntax $\Ff_\app=F_\app\Ff_\deco$
is recovered by dropping the decorations. 
The explicit syntax $\Ff_\expl=F_\expl\Ff_\deco$ 
is the theory in the explicit logic generated by 
a type $P_i$  
and a term $\ti{\ft_i}_\ppg : P_i\to E$ for each $i\in\Const$, 
which form a coproduct $(\ti{\ft_i}_\ppg :P_i\to E)_{i\in\Const}$.
So, for each $i$, the operation $\ti{\fc_i}:E \to P_i+E $
is defined up to strong equations by the weak equations: 
  $$ \begin{cases}
  \ti{\fc_i} \circ \ti{\ft_i}_\ppg \eqw \inl_{P_i} & : P_i\to P_i+E \\ 
  \ti{\fc_i} \circ \ti{\ft_j}_\ppg \eqw \inr_{P_i} \circ \ti{\ft_j}_\ppg 
   & : P_j\to P_i+E \; \mbox{ for each }j\ne i \\
   \end{cases} $$

The explicit theory $\Tt_\expl$ is made of the category of sets 
with the equality as congruence, 
with a distinguished set $\Exc$ called the set of exceptions, 
with disjoint unions with $\Exc$, 
and with a coproduct on $\Const$
with vertex $\Exc$, denoted by $(\sft_i:\Par_i\to \Exc)_{i\in\Const}$,  
so that $\Exc=\sum_{j\in\Const} \Par_j$.  
The explicit semantics $M_\expl:\Ff_\expl\to\Tt_\expl$ is the model 
(in the explicit logic) which maps 
$E$ to $\Exc$ and, for each $i\in\Const$,   
the type $P_i$ to the set $\Par_i$
and the operations $\ft_i$ and $\fc_i$
to the functions $\sft_i$ and $\sfc_i$, respectively. 
The decorated semantics $M_\deco:\Ff_\deco\to\Tt_\deco$ 
is obtained from the explicit semantics $M_\expl:\Ff_\expl\to\Tt_\expl$
thanks to the adjunction $F_\expl\dashv G_\expl$. 

The next result is dual to proposition~\ref{proposition:states-prop}, 
it can be proved in the dual way, using the decorated logic
for exceptions. 
It is the key lemma for proving proposition~\ref{proposition:exceptions-prop},
which says that catching an exception of constructor $i$ 
by throwing the same exception is like doing nothing. 

\begin{prop}
\label{proposition:exceptions-key-prop}
For every $i\in\Const$:
$$ \begin{cases}
 \ft_i^\ppg \circ \fc_i^\ctc \eqs \id_\empt^\pure  & 
  \mbox{ in the decorated logic } \\ 
 \ti{\ft_i} \circ \ti{\fc_i} \eqs \id_E & 
  \mbox{ in the explicit logic } \\ 
\end{cases} $$
\end{prop}

\subsection{Extending the decorated logic}
\label{subsec:exceptions-logic}

In the previous section~\ref{subsec:exceptions-dual} 
the key operations $\ft_i^\ppg$'s and $\fc_i^\ctc$'s have been defined;
in the next section~\ref{subsec:exceptions-encaps} 
they will be used for building the decorated raising and handling operations.
For this purpose, some rules must be added to
the decorated logic for exceptions; this is done now. 

\begin{defi}
\label{definition:exceptions-encaps-rule}
The decorated logic for exceptions $\catT_{\deco,\exc}$ 
is extended as $\catT_{\deco,\exc}^+$ by adding the following rules. 
\begin{itemize}
\item 
For each point $X$ there is a \emph{decorated sum} $X=X+\empt$,
in the sense that for each propagator $g^\ppg:X\to Y$ and 
each catcher $k^\ctc:\empt\to Y$
there is a catcher $\cotuple{g\,|\,k}^\ctc:X\to Y$,
unique up to strong equations, such that 
$\cotuple{g\,|\,k}^\ctc \eqw g^\ppg$ and 
$\cotuple{g\,|\,k}^\ctc \circ \cotu_X^\pure \eqs k^\ctc$.
$$  \xymatrix@C=10pc@R=1.5pc{
  X \ar[r]^(.3){\cotuple{g\,|\,k}^\ctc} \ar@<1ex>@/^3ex/[r]^{g^\ppg}_{\eqw} & Y \\
  \empt \ar[u]^{\cotu^\pure} \ar[ru]_{k^\ctc}^{\eqs} & \\
  } $$
In addition, whenever $f^\ppg:\empt\to Y$ is a propagator 
(which implies that $f^\ppg \eqs \cotu_Y^\pure$) 
then $\cotuple{g\,|\,f}:X\to Y$ is a propagator
(so that the weak equation $\cotuple{g\,|\,f}^\ppg \eqw g^\ppg$
is strong: $\cotuple{g\,|\,f}^\ppg \eqs g^\ppg$).
In particular, we will use the fact that 
$\cotuple{g^\ppg\,|\,\cotu_Y^\pure} \eqs g^\ppg$.
$$  \xymatrix@C=10pc@R=1.5pc{
  X \ar[r]_{\cotuple{g\,|\,\cotu_Y}^\ppg} 
    \ar@<1ex>@/^3ex/[r]^(.3){g^\ppg}_{\eqs} & Y \\
  } $$
\item  
For each catcher $k^\ctc:X\to Y$ there is a propagator $\toppg{k}^\ppg:X\to Y$,
unique up to strong equations, such that $\toppg{k}^\ppg \eqw k^\ctc$.
$$  \xymatrix@C=10pc{
  X \ar[r]_{k^\ctc} \ar@<1ex>@/^3ex/[r]^(.3){\toppg{k}^\ppg}_{\eqw}  & Y \\ 
  } $$
Thus, whenever $f^\ppg:X\to Y$ is a propagator 
then $\toppg{f}^\ppg \eqs f^\ppg$.
$$  \xymatrix@C=10pc{
  X \ar[r]_{f^\ppg} \ar@<1ex>@/^3ex/[r]^(.3){\toppg{f}^\ppg}_{\eqs}  & Y \\ 
  } $$
\end{itemize}
\end{defi}

Let us check that $\catT_{\deco,\exc}$ can be replaced by 
$\catT_{\deco,\exc}^+$ in the span of logics $\zoom_{\deco}$:
we have to check that both morphisms $F_{\app,\deco}$ and $F_{\expl,\deco}$ 
map the new rules to rules in the apparent and in the explicit logic, 
respectively. 
This is obvious for the apparent logic, 
where $k\eqs\cotu_Y$, $\cotuple{g\,|\,k}\eqs g$ and $\toppg{f} \eqs f$. 
For the explicit logic, 
for each $\ti{g}_\ppg:X\to Y+E$ and $\ti{k}:E\to Y+E$ we have 
$\ti{\cotuple{g\,|\,k}} = \cotuple{\ti{g}_\ppg\,|\,\ti{k}} : X+E\to Y+E$. 
It follows that 
$\ti{\cotuple{g\,|\,\cotu_Y}} \eqs \ti{g}$, which propagates exceptions.
And for each $\ti{k}:X+E\to Y+E$ we have 
$\ti{\toppg{k}}_\ppg \eqs \ti{k}\circ\inl_X:X\to Y+E$.
  
\subsection{Encapsulating exceptions}
\label{subsec:exceptions-encaps}

This section is a ``decorated version'' 
of section~\ref{subsec:explicit-exceptions}. 
We show that the $\ft_i^\ppg$'s and $\fc_i^\ctc$'s are 
the key operations for dealing with exceptions in the decorated logic. 
More precisely, we prove that,
for each constructor $i$, the \emph{raise} operation
is built from pure operations and a unique propagator $t_i$,
and the \emph{handle} operation 
is built from propagators and a unique catcher~$c_i$. 
There are at least two reasons for not using $\ft_i$ and $\fc_i$ directly: 
firstly in a programming language there is usually no name 
and no intuition for the ``empty'' type $\empt$, 
secondly the handling of exceptions is a powerful programming technique
which must be carefully encapsulated:
while most operations are allowed to throw exceptions, 
only some very special operations are allowed to catch exceptions. 

First, let us focus on \emph{raising} exceptions. 
This operation is a propagator,
it calls the key thrower $\ft_i^\ppg :P_i \to \empt$
and ``hides'' the empty type by mapping it into the required type of results.

\begin{defi} 
\label{definition:exceptions-raise} 
For each $i$ in $\Const$ and each object $Y$, 
the propagator ``\emph{raise 
(or throw) an exception of constructor $i$ in $Y$}'' 
is 
$$ \rais{i}{Y}^\ppg = \throw{i}{Y}^\ppg = 
\cotu_Y^\pure \circ \ft_i^\ppg :P_i \to Y  $$  
  \begin{equation}
  \label{diag:raise-decorated} 
  \xymatrix@C=4pc@R=1.5pc{
  P_i \ar[rr]^{\rais{i}{Y}^\ppg} \ar@<-.5ex>[rrd]_{\ft_i^\ppg} && 
    Y \ar@{}[lld]|(.3){\eqs} \\ 
  && \empt \ar[u]_{\cotu^\pure}  \\ 
  } 
  \end{equation}
\end{defi}

Now, let us consider the \emph{handling} of exceptions,
which calls the key catchers $\fc_i^\modi$'s.
Let $f^\ppg:X\to Y$ be some propagator.  
For handling exceptions of constructors $i_1,\dots,i_n$ raised by $f$, 
using propagators $g_1^\ppg:\Par_{i_1}\to Y,\dots,g_n^\ppg:\Par_{i_n}\to Y$, 
the handling process builds a propagator: 
  $$ (\handlerec{f}{i_1}{g_1}{i_n}{g_n})^\ppg \;=\;
  (\try{f}{\catchrec{i_1}{g_1}{\catchrec{i_2}{g_2}{...\catch{i_n}{g_n}}}})^\ppg 
  $$ 
which is also denoted in a more compact way as
  $$ (\handlen{f}{(i_k\!\!\Rightarrow\!\! g_k)_{1\leq k\leq n}})^\ppg  \;=\; 
    (\try{f}{\catchn{i_k\{g_k\}_{1\leq k\leq n}}})^\ppg : X\to Y $$ 
    
\begin{defi} 
\label{definition:exceptions-handle} 
For each propagator $f^\ppg:X\to Y$, each $n\geq1$, 
each exception constructors $i_1,...,i_n$ and each 
propagators $g_1^\ppg:P_{i_1}\to Y,...,g_n^\ppg:P_{i_n}\to Y$, 
the propagator 
``\emph{handle the exception $e$ raised in $f$, if any,   
with $g_1$ if $e$ has constructor $i_1$, 
otherwise with $g_2$ if $e$ has constructor $i_2$, ..., 
otherwise with $g_n$ if $e$ has constructor $i_n$}'', 
is 
  $$ (\handlen{f}{(i_k\!\!\Rightarrow\!\! g_k)_{1\leq k\leq n}})^\ppg  \;=\; 
    (\try{f}{\catchn{i_k\{g_k\}_{1\leq k\leq n}}})^\ppg : X\to Y $$ 
defined as follows.
\begin{enumerate}
\item The catchers $(\catchn{i_k\{g_k\}_{p\leq k\leq n}})^\ctc : \empt\to Y$
are defined recursively by  
  $$ (\catchn{i_k\{g_k\}_{p\leq k\leq n}})^\ctc \;=\;
    \begin{cases} 
     \cotuple{g_p^\ppg \;|\; (\catchn{i_k\{g_k\}_{p+1\leq k\leq n}})^\ctc}^\ppg 
       \circ \fc_{i_p}^\ctc 
       & \mbox{ when } p<n \\
     g_n^\ppg  
       \circ \fc_{i_n}^\ctc  & \mbox{ when } p=n \\
   \end{cases} $$ 
  \begin{equation}
  \label{diag:handle-decorated-catch} 
  \xymatrix@C=3pc{
  \empt \ar[r]^{\fc_{i_{p+1}}^\ctc} & 
    P_i \ar[rrrr]^(.4){\cotuple{ g_p \;|\; \dots }^\ctc } &
    \ar@{}[dl]|(.4){\eqs} &&& Y \\ 
  & \empt \ar[u]^{\cotu^\pure} \ar[rrrru]_{\dots} &&&& \\
  } 
  \end{equation}
where $\dots$ stands 
for $(\catchn{i_k\{g_k\}_{p+1\leq k\leq n}})^\ctc $ when $p<n$ 
and for $\cotu_Y^\pure$ when $p=n$,
since $\cotuple{ g_n^\ppg \;|\; \cotu_Y^\pure }^\ppg \eqs  g_n^\ppg$.
\item Then the catcher $\Handle^\ctc:X\to Y$ is defined as 
$$  \Handle^\ctc \;=\; 
    \cotuple{ \id_Y^\pure \;|\; (\catchn{i_k\{g_k\}_{1\leq k\leq n} })^\ctc} 
      \circ f^\ppg : 
    X\to Y $$
  \begin{equation}
  \label{diag:handle-decorated-intermediate} 
  \xymatrix@C=3pc{
  X \ar[r]^{f^\ppg} & 
    Y \ar[rrrr]^(.4){\cotuple{ \id \;|\; \catchn{i_k\{g_k\}_{1\leq k\leq n} }}^\ctc} &
    \ar@{}[dl]|(.4){\eqs} &&& Y \\ 
  & \empt \ar[u]^{\cotu^\pure} \ar[rrrru]_{(\catchn{i_k\{g_k\}_{1\leq k\leq n}})^\ctc} &&&& \\
  } 
  \end{equation}  
\item Finally the handling function is the propagator 
$(\toppg\Handle)^\ppg$
$$ (\try{f}{\catchn{i_k\{g_k\}_{1\leq k\leq n}}})^\ppg \;=\; 
  (\toppg{ \Handle})^\ppg : X \to Y$$
  \begin{equation}
  \label{diag:handle-decorated-handle} 
  \xymatrix@C=16pc{
  X \ar[r]_{\Handle^\ctc} 
    \ar@<1ex>@/^3ex/[r]^{(\try{f}{\catchn{i_k\{g_k\}_{1\leq k\leq n}}})^\ppg}_{\eqw}  & Y \\ 
  } 
  \end{equation}
\end{enumerate}
\end{defi}

Altogether, we get:
  $$ \xymatrix@C=2pc@R=1pc{
  X \ar[rr]^(.6){f^\ppg} 
    \ar@<1ex>@/^4ex/[rrrrrrr]^{(\handlen{f}{(i_k \Rightarrow g_k)_{1\leq k\leq n}})^\ppg}_{\eqw} && 
    Y \ar[rrrrr]|(.4){\;[\id|\dots]^\ctc\;} &&&&& Y \\ 
  && \empt \ar[u]^{\cotu^\pure} \ar[r]^{\fc_{i_1}^\ctc} & 
    P_{i_1}  \ar@{}[u]|{\eqs} \ar@/_/[rrrru]|(.4){\;\cotuple{g_1\,|\,\dots}^\ctc\;} 
      &&&& \\
  &&& \empt \ar[u]^{\cotu^\pure} \ar[r]^{\fc_{i_2}^\ctc} \ar@{..}[ddr] & 
    P_{i_2}  \ar@{}[u]|{\eqs} \ar@/_/[rrruu]|(.4){\;\cotuple{g_2\,|\,\dots}^\ctc\;} 
      \ar@{..}[ddr] &&& \\ 
  \\ 
  &&&& \empt \ar[r]^{\fc_{i_{n-1}}^\ctc} & 
    P_{i_{n-1}} \ar@/_/[rruuuu]|(.4){\;\cotuple{g_{n-1}\,|\,\dots}^\ctc\;} && \\ 
  &&&&& \empt \ar[u]^{\cotu^\pure} \ar[r]^{\fc_{i_n}^\ctc} & 
    P_{i_n}  \ar@{}[u]|(.8){\eqs} \ar@/_/[ruuuuu]|(.4){\;g_n^\ppg\;} & \\  
  }$$
When $n=1$, this becomes simply:
$$ (\handle{f}{i}{g})^\ppg = (\try{f}{\catch{i}{g}})^\ppg = 
  \toppg{([\id_Y \,|\, g\circ \fc_i] \circ f)} : X\to Y $$ 
  $$ \xymatrix@C=5pc{ 
  X \ar[r]_(.6){f^\ppg} \ar@<1ex>@/^3ex/[rrr]^{(\handlen{f}{i\Rightarrow g})^\ppg}_{\eqw} & 
    Y \ar[rr]_{[\id\,|\,g\circ \fc_i]^\ctc} && Y \\ 
  & \empt \ar[u]_{\cotu^\pure} \ar[r]_{\fc_i^\ctc} & 
    P_i \ar@/_/[ru]_{g^\ppg} \ar@{}[u]|{\eqs} & \\
  } $$

It is easy to check that by applying the
expansion morphism $F_\expl$ to the decorated definitions
of \emph{raise} and \emph{handle} 
we get the explicit description from section~\ref{subsec:explicit-exceptions}:
diagrams~\ref{diag:raise-explicit}, \ref{diag:handle-explicit-catch},
\ref{diag:handle-explicit-intermediate}, \ref{diag:handle-explicit-handle},
are mapped respectively to 
diagrams~\ref{diag:raise-decorated}, \ref{diag:handle-decorated-catch},
\ref{diag:handle-decorated-intermediate}, \ref{diag:handle-decorated-handle}. 
According to claim~\ref{claim:exceptions-explicit}, 
the explicit language $\Lang_{\expl,\exc}$ 
provides the intended semantics of exceptions. 
By adjunction (remark~\ref{remark:dialog-effects-expl}) 
this is also the semantics 
of the language with effect $\Lang_{\deco,\exc}$, 
hence the following result.

\begin{prop}
\label{proposition:exceptions} 
The language with effect $\Lang_{\deco,\exc}$ 
provides the intended semantics of exceptions. 
\end{prop}

\subsection{Some properties of exceptions} 
\label{subsec:exceptions-prop}

The next proposition 
shows that catching an exception of constructor $i$ 
by throwing the same exception is like doing nothing. 
Indeed, by expansion this result implies that 
in the semantics (as in section~\ref{subsec:explicit-exceptions})
for all $e\in\Exc\,$,  
$ (\catch{i}{\throw{i}{Y}})(e) = e \in \Exc \subseteq Y+\Exc$. 

\begin{prop}
\label{proposition:exceptions-prop}
For every $i\in\Const$, in the decorated logic: 
$$ (\catch{i}{\throw{i}{Y}})^\ctc \eqs \cotu_Y^\pure $$
\end{prop}

\proof 
Let $g= \throw{i}{Y} :P_i \to Y $. 
By definition~\ref{definition:exceptions-handle} we have 
$\catch{i}{g} = \cotuple{g|\cotu_Y} \circ \fc_i : \empt \to Y $, 
and since $g$ is a propagator,
by definition~\ref{definition:exceptions-encaps-rule}  
we have $\cotuple{ g| \cotu_Y }\eqs g$, 
so that $\catch{i}{g} \eqs g \circ \fc_i$. 
By definition~\ref{definition:exceptions-raise} we have 
$\throw{i}{Y} = \cotu_Y \circ \ft_i :P_i \to Y $
hence 
$\catch{i}{g} \eqs \cotu_Y \circ \ft_i \circ \fc_i$,
and since $ \ft_i \circ \fc_i \eqs \id_\empt $
by proposition~\ref{proposition:exceptions-key-prop} 
we get $\catch{i}{\throw{i}{Y}} \eqs \cotu_Y$. 
 $$
\xymatrix@C=10pc@R=3pc{
&& Y \ar[d]^{\cotu^\pure} \\
\empt \ar[r]_{\fc_i^\ctc} \ar@/_5ex/[rr]_{(\catch{i}{\throw{i}{Y}})^\ctc}^{=} 
  \ar@/^3ex/[rru]^{\id^\pure}_{\eqs} & 
    P_i \ar[r]_(.4){\cotuple{\throw{i}{Y}|\cotu}^\ppg} 
      \ar@/^3ex/[r]^(.6){\throw{i}{Y}^\ppg}_{\eqs} 
      \ar@/^2ex/[ru]^(.4){\ft_i^\ppg}_(.6){=} & 
  Y \\ 
}$$
\qed

\begin{rem} 
The three propagators  
$$ \begin{cases}
   \try{f}{\catchrec{i}{g}{\catch{j}{h}}} \\
   \try{\try{f}{\catch{i}{g}}}{\catch{j}{h}} \\
   \try{f}{\catch{i}{\try{g}{\catch{j}{h}}}}  \\
   \end{cases}  $$
do not behave in the same way:
whenever $f(x)$ raises an exception $\ft_i(a)$ of constructor $i$
and $g(a)$ raises an exception $\ft_j(b)$ of constructor $j$,
the first propagator returns $\ft_j(b)$ (uncaught) 
while the second and the third ones return $h(b)$;
whenever $f(x)$ raises an exception $\ft_j(b)$ of constructor $j$, 
the first and the second propagators return $h(b)$ 
while the third one returns $\ft_j(b)$ (uncaught). 
The differences can be seen from the diagrams. 
  $$ 
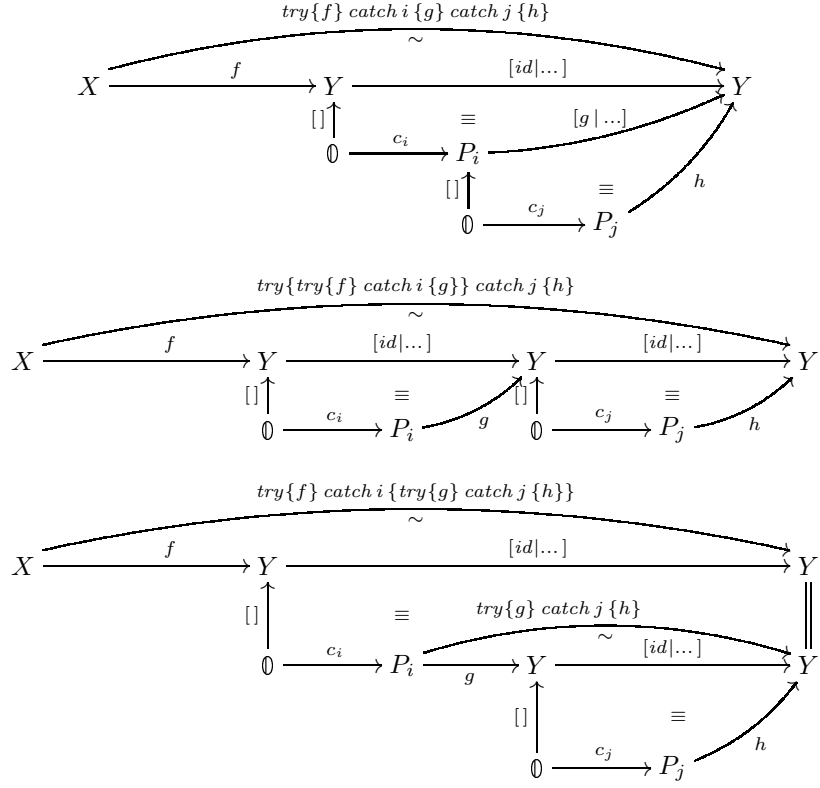
\xymatrix@C=3pc@R=1pc{
  X \ar[rr]^(.6){f} \ar@<1ex>@/^4ex/[rrrrr]^{\try{f}{\catchrec{i}{g}{\catch{j}{h}}}}_{\eqw} && 
    Y \ar[rrr]^{[\id|\dots]} &&& Y \\ 
  && \empt \ar[u]^{\cotu} \ar[r]^{\fc_i} & 
    P_i \ar@{}[u]|{\eqs} \ar@/_/[rru]^{\cotuple{g\,|\,\dots}} && \\ 
  &&& \empt \ar[u]^{\cotu} \ar[r]^{\fc_j} & 
    P_j \ar@{}[u]|{\eqs} \ar@/_/[ruu]_{h} & \\
  } $$
 \smallskip
 $$  \xymatrix@C=3pc@R=1pc{
  X \ar[rr]^(.6){f} \ar@<1ex>@/^4ex/[rrrrrr]^{\try{\try{f}{\catch{i}{g}}}{\catch{j}{h}}}_{\eqw} && 
    Y \ar[rr]^{[\id|\dots]} && Y \ar[rr]^{[\id|\dots]} && Y \\ 
  && \empt \ar[u]^{\cotu} \ar[r]^{\fc_i} & 
    P_i \ar@{}[u]|{\eqs} \ar@/_/[ru]_{g} &
    \empt \ar[u]^{\cotu} \ar[r]^{\fc_j} & 
    P_j \ar@{}[u]|{\eqs} \ar@/_/[ru]_{h} & \\
  }$$
 \smallskip
 $$  \xymatrix@C=3pc@R=2pc{
  X \ar[rr]^(.6){f} \ar@<1ex>@/^4ex/[rrrrrr]^{\try{f}{\catch{i}{\try{g}{\catch{j}{h}}}}}_{\eqw} && 
    Y \ar[rrrr]^{[\id|\dots]} &&&& Y \\ 
  && \empt \ar[u]^{\cotu} \ar[r]^{\fc_i} & 
    P_i \ar@{}[u]|{\eqs} \ar[r]_{g} \ar@<.5ex>@/^3ex/[rrr]^(.4){\try{g}{\catch{j}{h}}}_{\eqw} & 
    Y \ar[rr]^{[\id|\dots]} && Y \ar@{=}[u] \\ 
  &&&& \empt \ar[u]^{\cotu} \ar[r]^{\fc_j} & P_j \ar@<-.5ex>@{}[u]|{\eqs} \ar@/_/[ru]_{h}& \\
  }$$
\end{rem}

The next result is proved in appendix~\ref{app:exceptions}.

\begin{prop}
\label{proposition:exceptions-two}
For every $i,j\in\Const$, in the decorated logic: 
\renewcommand{\arraystretch}{1.5}
$$ \begin{array}{l}
\try{f}{\catchrec{i}{g}{\catch{j}{h}}} \eqs \try{f}{\catchrec{j}{h}{\catch{i}{g}}}
\mbox{ if } i\ne j \\ 
\try{f}{\catchrec{i}{g}{\catch{i}{h}}} \eqs \try{f}{\catch{i}{g}} \\ 
\end{array}$$
\renewcommand{\arraystretch}{1}
\end{prop}

\begin{rem}
\label{remark:exceptions-catchall}
The \emph{catch} construction is easily extended to 
a \emph{catch-all} construction like \texttt{catch(...)} in \cpp. 
We add to the decorated logic for exceptions 
a pure unit type $\unit$,
which means, a type $\unit$ such that 
for each type $X$ there is a pure term $\tu_X:X\to\unit$,
unique up to strong equations. 
Then we add a catcher $\fc_\all^\ctc:\empt\to \unit$ 
with the equations $\fc_\all \circ \ft_j \eqw \tu_{P_j} $ 
for every $j\in\Const$,
which means that $\fc_\all$ catches exceptions of the form $\ft_j(a)$ 
for every $j$ and forgets the value $a$. 
For each propagators $f^\ppg:X\to Y$ and $g^\ppg:\unit\to Y$, 
the propagator 
``\emph{handle the exception $e$ raised in $f$, if any, with $g$}''
is 
$$ (\handle{f}{\all}{g})^\ppg = 
  \toppg{([\id_Y \,|\, g\circ \fc_\all] \circ f)} : X\to Y $$ 
  $$ \xymatrix@C=5pc{ 
  X \ar[r]_(.6){f^\ppg} \ar@<1ex>@/^3ex/[rrr]^{(\handlen{f}{\all\Rightarrow g})^\ppg}_{\eqw} & 
    Y \ar[rr]_{[\id \,|\, g\circ \fc_\all]^\ctc} && Y \\ 
  & \empt \ar[u]_{\cotu} \ar[r]_{\fc_\all^\ctc} & 
    \unit \ar@/_/[ru]_{g^\ppg} \ar@{}[u]|{\eqs} & \\
  } $$
The semantics of the \emph{catch-all} construction is easily derived 
from this diagram, 
as a function $(\handle{f}{\all}{g}):X\to Y+\Exc$ 
where $\Exc=\sum_{j\in\Const}Par_j$
and $g:\unit\to Y+\Exc$ is a constant:  
\begin{tabbing}
 9999 \= 9999 \= 9999 \= 9999 \= \kill 
 \> For each $x\in X+\Exc$, $\;(\handle{f}{\all}{g})(x) \in Y+\Exc$ 
   is defined by: \\
 \>\> if $x\in \Exc$ then return $x\in \Exc \subseteq Y+\Exc$; \\
 \>\> // \textit{now $x$ is not an exception} \\ 
 \>\> compute $y:=f(x) \in Y+\Exc$; \\
 \>\> if $y\in Y$ then return $y\in Y \subseteq Y+\Exc$; \\
 \>\> // \textit{now $y$ is an exception} \\ 
 \>\> return $g \in Y+\Exc$. 
 \end{tabbing}
This is indeed the required semantics of the ``catch-all'' construction.
It may be combined with other catchers,
and it follows from this construction that 
every catcher following a ``catch-all'' 
is syntactically allowed, but never executed. 
\end{rem}

\section{The duality}
\label{sec:dual}

The previous results are summarized in section~\ref{subsec:dual-effects}, 
then some remarks about other semantical issues are 
outlined in section~\ref{subsec:dual-other}.

\subsection{Duality of states and exceptions as effects}  
\label{subsec:dual-effects}

Given a set $I$,
let $\zoom_{\deco,\st}$ be the span of diagrammatic logics for states 
with respect to the set of locations $I$ 
as defined in section~\ref{subsec:states-zoom}.
In this span, 
let $\Lang_{\deco,\st}$ be the language with effects for states 
as defined in section~\ref{subsec:states-effect}. 
Then proposition~\ref{proposition:states}
states that $\Lang_{\deco,\st}$ provides the intended 
semantics of states.

Given a set $I$,
let $\zoom_{\deco,\exc}$ be the span of diagrammatic logics for exceptions 
with respect to the set of exceptions constructors $I$ 
as defined in section~\ref{subsec:exceptions-dual}. 
In this span, 
let $\Lang_{\deco,\exc}$ be the language with effects for exceptions 
as defined in section~\ref{subsec:exceptions-dual}. 
Then proposition~\ref{proposition:exceptions}
states that $\Lang_{\deco,\exc}$ provides the intended semantics of exceptions.
It should be reminded that the
whole process of raising and handling exceptions 
does rely on the key functions $\ft_i$ and $\fc_i$: 
this has been checked in section~\ref{subsec:exceptions-encaps}. 

Figure~\ref{fig:duality-syntax} recapitulates the properties of 
the functions \emph{lookup} ($\fl_i$) and \emph{update} ($\fu_i$) 
for states on the left, 
and the properties of the functions \emph{key throw} ($\ft_i$) and 
\emph{key catch} ($\fc_i$) for exceptions on the right. 
By expansion, figure~\ref{fig:duality-syntax} 
gives rise to figure~\ref{fig:duality-semantics}.
Our main result (theorem~\ref{theorem:duality}) follows immediately; 
it means that  
the well-known duality between categorical products and coproducts 
can be extended as a duality 
between the lookup and update functions for states on one side 
and the key throwing and catching functions for exceptions on the other.
The notion of opposite categories and duality is 
extended in the straightforward way to diagrammatic logics 
and to spans of diagrammatic logics.

\begin{figure}
\renewcommand{\arraystretch}{1.3}
$$ \begin{array}{|c|c|}
\hline
\makebox[60mm]{\textbf{States}} & 
\makebox[60mm]{\textbf{Exceptions}} \\ 
\hline
i\in\Loc,\; \Val_i,\; &
i\in\Const,\; \Par_i,\; \\
\unit \mbox{ terminal } &
\empt \mbox{ initial } \\
\tu_i^\pure : \Val_i \to \unit & 
\empt \lto \Par_i :  \cotu_i^\pure \\ 
\hline
\fl_i^\acc : \unit\to\Val_i &
\empt\lto\Par_i : \ft_i^\ppg \\ 
\fu_i^\modi : \Val_i\to \unit &
\Par_i\lto \empt : \fc_i^\ctc \\ 
\hline
\xymatrix@R=1pc@C=3pc{
\Val_i \ar[r]^{\id} \ar[d]_{\fu_i} & \Val_i \ar[d]^{\id} \\
\unit \ar[r]^{\fl_i} & \Val_i \ar@{}[ul]|{\eqw} \\
} &
\xymatrix@R=1pc@C=3pc{
\Par_i & \Par_i \ar[l]_{\id} \\
\empt \ar[u]^{\fc_i} & \Par_i \ar[l]_{\ft_i} \ar[u]_{\id} \ar@{}[ul]|{\eqw} \\
} \\
\xymatrix@R=1pc@C=2pc{
\Val_i \ar[r]^{\tu} \ar[d]_{\fu_i} & 
  \unit \ar[r]^{\fl_j} & \Val_j \ar[d]^{\id} \\
\unit \ar[rr]^{\fl_j} && \Val_j \ar@{}[ull]|{\eqw} \\
} & 
\xymatrix@R=1pc@C=2pc{
\Par_i & \empt \ar[l]_{\cotu}
  & \Par_j \ar[l]_{\ft_j} \\
\empt \ar[u]^{\fc_i} && \Par_j \ar[ll]_{\ft_j} \ar[u]_{\id} \ar@{}[ull]|{\eqw} \\} \\ 
(j\ne i) & (j\ne i) \\
\hline
\end{array} $$
\renewcommand{\arraystretch}{1}
\caption{Duality of decorated syntax}
\label{fig:duality-syntax}
\end{figure}

\begin{thm}
\label{theorem:duality} 
With the previous notations, 
the span of diagrammatic logics $\zoom_\exc$ for exceptions 
is opposite to the span of diagrammatic logics $\zoom_\st$ for states 
and the language with effects $\Tt_{\deco,\exc}$ for exceptions 
is dual to the language with effects $\Tt_{\deco,\st}$ for states.
\end{thm}

\subsection{Other semantics}
\label{subsec:dual-other}

Equations~(\ref{eq:exceptions}) relating the key throw and catch  
operations may be oriented from left to right in order to get the 
usual \emph{operational semantics} of exceptions: 
when an exception is thrown by some occurrence of $\ft_i$, 
the execution jumps to the first occurrence of $\fc_i$ and 
wipes out the pair $(\ft_i,\fc_i)$ and everything between them. 

In a dual way, equations~(\ref{eq:states}) relating the lookup and update 
operations may be oriented from left to right,
but this does not provide the usual operational semantics of states.
In fact, equations~(\ref{eq:states}) are related 
to the \emph{Hoare-Floyd semantics} of states:
they give rise to the basic occurrences of the assignment axiom 
$ \{G[e/X]\} \; X:=e \; \{G\} $, namely: 
  $$ \{e=n\} \; X:=e \; \{X=n\} \;\mbox{ and }\; 
  \{Y=n\} \; X:=e \; \{Y=n\} \;\mbox{ when }\; Y\ne X $$
>From the decorated point of view, 
the value $n$ is pure, 
the expressions $e$, $e=n$, $X=n$ and $Y=n$ are accessors, 
the command $X:=e$ is a modifier
and the equalities are weak equations. 
The axioms mean that whenever $e^\acc \eqw n^\pure$
then $\fl_X^\acc \circ \fu_X^\modi \circ e^\acc \eqw n^\pure$ 
and $\fl_Y^\acc \circ \fu_X^\modi \circ e^\acc \eqw \fl_Y^\acc $ if $Y\ne X$, 
which is easily derived from equations~\ref{eq:states}. 

\section*{Conclusion}

We have discovered a symmetry between the key notions 
underlying the effects of states and exceptions, 
thanks to our approach of computational effects 
relying on spans of diagrammatic logics. 
A consequence is that the duality principle can be applied for 
deriving properties of exceptions
from the properties of states. 
Another consequence is that 
this symmetry provides a new point of view on exceptions, 
mainly by distinguishing the key catching operation 
from the surrounding conditionals in the handling process. 

This symmetry between states and exceptions is deeply hidden, 
which may explain that our result is,
as far as we know, completely new. 
First, as seen in the paper, 
for states the key operations are visible, 
while for exceptions they are encapsulated. 
In addition, most features which we might  
want to add will contribute to hide the duality:
this happens for instance simply when adding pure constants 
$a^\pure:\unit\to V_i$ for states and $a^\pure:\unit\to P_i$ 
for exceptions, not $a^\pure: P_i\to\empt$.
Adding products on one side and coproducts on the other, 
as in appendix~\ref{app}, preserves the duality.
Adding both products and coproducts on either side preserves the duality, 
but the distributivity or extensivity property, which 
is usually assumed, does not preserve it. 
Adding exponentials in order to get a lambda-calculus 
would be desirable, but this might further obscure the duality.  
Many questions are still open, for instance 
about a similar duality applying for other effects,
or about the combination of effects from this point of view.

\bibliographystyle{alpha}

\appendix
\section{}
\label{app}
\renewcommand{\arraystretch}{1.5}

In this appendix we consider two equations for states:  
equations $(6)$ and $(3)$ in the list of equations~\ref{eq:states-seven}
(section~\ref{subsec:explicit-states}),      
and the dual equations for exceptions, 
all of them in the decorated logic. 
In the decorated proofs below, 
the associativity and identity rules are skipped  
and the decoration of morphisms is often omitted.

\subsection{States}
\label{app:states}

Equations $(6)$ and $(3)$ in the list~\ref{eq:states-seven} 
(section~\ref{subsec:explicit-states}) are:
$$ \begin{array}{ll}
(6) 
& \forall i\ne j\in\Loc ,\; 
\forall\,s\in\St ,\; a\in\Val_i ,\; b\in\Val_j ,\;\; 
\sfu_j(b,\sfu_i(a,s)) = \sfu_i(a,\sfu_j(b,s)) \in\St \\ 
(3) 
& \forall i\in\Loc ,\; \forall\,s\in\St ,\; a,a'\in\Val_i ,\;\; 
\sfu_i(a',\sfu_i(a,s)) = \sfu_i(a',s) \in\St \\ 
\end{array} $$
Since these equations have two values as arguments,
we will use the notion of semi-pure product from \cite{DDR11}.  
In the decorated logic for states, 
the \emph{product} of two objects $A$ and $B$ 
is an object $A\times B$ with two pure morphisms: the projections
$\pi_1^\pure:A\times B \to A$ and $\pi_2^\pure:A\times B \to B$, 
which satisfy the usual categorical product property 
with respect to the pure morphisms
(so that, as usual, the projections 
$\pi_1^\pure:A\times \unit \to A$ and $\pi_2^\pure:\unit\times B \to B$
are isomorphisms).
The \emph{product} of two pure morphisms $f^\pure:A\to C$ 
and $g^\pure:B\to D$ is a pure morphism 
$(f\times g)^\pure:A\times B\to C\times D$ 
which is characterized, up to $\eqs$, by: 
\begin{align*}
\pi_1^\pure \circ (f \times g)^\pure & \eqs  f^\pure \circ \pi_1^\pure \\
\pi_2^\pure \circ (f \times g)^\pure & \eqs  g^\pure \circ \pi_2^\pure \\ 
\end{align*}
Such a property, symmetric in $f$ and $g$, 
cannot be satisfied by modifiers: 
indeed, the effect of building a pair of modifiers depends 
on the evaluation strategy. 
However, in \cite{DDR11} we define the \emph{semi-pure product} 
of a pure morphism $f^\pure:A\to C$ and a modifier $g^\modi:B\to D$,
as a modifier $(f\ltimes g)^\modi:A\times B\to C\times D$ 
which is characterized, up to $\eqs$,  
by the following decorated version of the product property: 
\begin{align*}
\tag*{\anprodpure}
\pi_1^\pure \circ (f \ltimes g)^\modi & \eqw f^\pure \circ \pi_1^\pure \\
\tag*{\anprodmodi}
\pi_2^\pure \circ (f \ltimes g)^\modi & \eqs g^\modi \circ \pi_2^\pure \\ 
\end{align*}
$$ 
\xymatrix@C=4pc{
  A \ar[r]^{f^\pure} & C \\
  A \times B \ar[r]^{(f \ltimes g)^\modi} \ar[u]^{\pi_1^\pure} \ar[d]_{\pi_2^\pure}
  \ar@{}[ur]|{\eqw}
   & C\times D \ar[u]_{\pi_1^\pure} \ar[d]^{\pi_2^\pure} \\
  B \ar[r]_{g^\modi} \ar@{}[ur]|{\eqs} & D \\ 
}
$$ 
The weak equations~\ref{eq:states} relating the 
functions $(\fu_i^\modi)_{i\in \Loc}$ and $(\fl_i^\acc)_{i\in\Loc}$
will be used as axioms in the proof trees with the following labels:
  \begin{align*}
  \fl_i^\acc \circ \fu_i^\modi & \eqw \id_{V_i}^\pure : V_i\to V_i
  \tag*{\anliui} \\
  \fl_j^\acc \circ \fu_i^\modi & \eqw \fl_j^\acc \circ \tu_{V_i}^\pure : V_i\to V_j 
    \; \mbox{ for each }j\ne i \tag*{\anliuj}\\
  \end{align*}

\medskip %
Equation $(6)$ is expressed in the decorated logic as:
\[ (6)_\st \; \forall i\neq j \in \Loc,\; 
  u_j^\modi \circ \pi_2^\pure \circ (u_i\rtimes \id_{V_j})^\modi \eqs 
    u_i^\modi \circ \pi_1^\pure \circ (\id_{V_i} \ltimes u_j)^\modi \]
This strong equation is equivalent to the family of weak equations:
\[ (6)_{\st,\ob} \; 
 \forall k\in\Loc, \forall i \neq j\in\Loc,\; 
  l_k^\acc \circ u_j^\modi \circ \pi_2^\pure \circ (u_i\rtimes \id_{V_j})^\modi \eqw 
    l_k^\acc \circ u_i^\modi \circ \pi_1^\pure \circ (\id_{V_i} \ltimes u_j)^\modi \] 
So, let $i,j,k\in\Loc$ with $i\ne j$. 

\begin{enumerate}

\item 
For $k\neq i,j$, the weak equation  
  \[ l_k\circ u_j\circ \pi_2 \circ (u_i\rtimes \id_{V_j}) \eqw 
  l_k \circ \tu_{V_i\times V_j}. \]
is proven in Figure~\ref{prooftree1} (proof \pnd). 
A symmetric proof shows that 
  \[l_k\circ u_i\circ \pi_1 \circ (\id_{V_i} \ltimes u_j) \eqw 
  l_k \circ \tu_{V_i\times V_j} \]
With the symmetry and transitivity of $\eqw$ 
this concludes the proof of equations~$(6)_{\st,\ob}$ 
when $k\neq i,j$.

\item 
When $k=i$, the weak equations 
$$ \begin{array}{l}
l_i\circ u_j\circ \pi_2\circ (u_i\rtimes \id_{V_j}) \eqw \pi_1 \\ 
l_i\circ u_i\circ \pi_1\circ (\id_{V_i}\ltimes u_j) \eqw \pi_1 \\ 
\end{array}$$
are proven in Figure~\ref{prooftree2} (proofs \png and \pnh). 
With the symmetry and transitivity of $\eqw$ 
this concludes the proof of equations~$(6)_{\st,\ob}$ 
when $k=i$.
The proof when $k=j$ is symmetric.
                    
\end{enumerate}

\begin{figure}[!h]
\begin{small}

\vspace{20pt}

\flushleft{Proof \pna:} 
\begin{prooftree}
\AxiomC{\anliuj}
\noLine
\UnaryInfC{$l_k \circ u_j \eqw l_k \circ \tu_{V_j}$}
\LeftLabel{\rnwsubs}
\UnaryInfC{$l_k\circ u_j\circ \pi_2\circ (u_i\rtimes \id_{V_j}) 
  \eqw l_k \circ \tu_{V_j} \circ \pi_2 \circ (u_i \rtimes \id_{V_j})$}
\end{prooftree}

\vspace{20pt}

\flushleft{Proof \pnb:} 
\begin{prooftree}
\AxiomC{$\tu_{V_j} \circ \pi_2: \unit\times V_j \to \unit$}
\LeftLabel{\rnwfinal}
\UnaryInfC{$\tu_{V_j} \circ \pi_2 \eqw \tu_{\unit\times V_j}$}
    \AxiomC{$\pi_1 : \unit\times V_j \to \unit$}
    \RightLabel{\rnwfinal}
    \UnaryInfC{${\pi_1} \eqw \tu_{\unit\times V_j}$}
    \RightLabel{\rnwsym}
    \UnaryInfC{$\tu_{\unit\times V_j} \eqw {\pi_1}$}
    \LeftLabel{\rnwtrans}
  \BinaryInfC{$\tu_{V_j}^\pure \circ \pi_2^\pure \eqw {\pi_1}^\pure$}
  \LeftLabel{\rnpa}
  \UnaryInfC{$\tu_{V_j}^\acc \circ \pi_2^\acc \eqw {\pi_1}^\acc$}
  \LeftLabel{\rnws}
  \UnaryInfC{$\tu_{V_j} \circ \pi_2 \eqs \pi_1$}
  \RightLabel{\rnssubs}
  \UnaryInfC{$\tu_{V_j} \circ \pi_2 \circ (u_i \rtimes \id_{V_j}) 
    \eqs \pi_1 \circ (u_i \rtimes \id_{V_j}) $}
    \AxiomC{\anprodmodi}
    \noLine
    \UnaryInfC{$\pi_1 \circ (u_i \rtimes \id_{V_j}) \eqs u_i \circ \pi_1$}
  \RightLabel{\rnstrans}
  \BinaryInfC{$\tu_{V_j} \circ \pi_2 \circ (u_i \rtimes \id_{V_j}) 
    \eqs u_i \circ \pi_1$}
  \LeftLabel{\rnsrepl}
  \UnaryInfC{$l_k \circ \tu_{V_j} \circ \pi_2 \circ (u_i \rtimes \id_{V_j}) 
    \eqs l_k \circ u_i \circ \pi_1$}
  \LeftLabel{\rnsw}
  \UnaryInfC{$l_k \circ \tu_{V_j} \circ \pi_2 \circ (u_i \rtimes \id_{V_j}) 
    \eqw l_k \circ u_i \circ \pi_1$}
\end{prooftree}

\vspace{20pt}

\flushleft{Proof \pnc:} 
\begin{prooftree}
\AxiomC{${\tu_{Vi}} \circ {\pi_1}: V_i\times V_j \to \unit$}
\LeftLabel{\rnwfinal}
\UnaryInfC{${\tu_{Vi}}^\pure \circ {\pi_1}^\pure \eqw \tu_{V_i \times V_j}^\pure$}
\LeftLabel{\rnpa}
\UnaryInfC{${\tu_{Vi}}^\acc \circ {\pi_1}^\acc \eqw \tu_{V_i \times V_j}^\acc$}
\LeftLabel{\rnws}
\UnaryInfC{$\tu_{Vi} \circ \pi_1 \eqs \tu_{V_i \times V_j}$}
\LeftLabel{\rnssubs}
\UnaryInfC{$l_k \circ \tu_{V_i} \circ \pi_1 \eqs l_k \circ \tu_{V_i\times V_j}$}
\LeftLabel{\rnsw}
\UnaryInfC{$l_k \circ \tu_{V_i} \circ \pi_1 \eqw l_k \circ \tu_{V_i\times V_j}$}
    \AxiomC{\anliuj}
    \noLine
    \UnaryInfC{$l_k \circ u_i \eqw l_k \circ \tu_{V_i}$}
    \RightLabel{\rnwsubs}
    \UnaryInfC{$l_k \circ u_i \circ \pi_1 \eqw l_k \circ \tu_{V_i} \circ \pi_1$}
  \LeftLabel{\rnwtrans}
  \BinaryInfC{$l_k \circ u_i \circ \pi_1 \eqw l_k \circ \tu_{Vi\times V_j}$}
\end{prooftree}

\vspace{20pt}

\flushleft{Proof \pnd:} 
\begin{prooftree}
\AxiomC{\pna}
    \AxiomC{\pnb}
  \LeftLabel{\rnwtrans}
  \BinaryInfC{$ l_k\circ u_j\circ \pi_2\circ (u_i\rtimes \id_{V_j}) 
    \eqw l_k \circ u_i \circ \pi_1 $}
      \AxiomC{\pnc}
    \LeftLabel{\rnwtrans}
    \BinaryInfC{$ l_k\circ u_j\circ \pi_2\circ (u_i\rtimes \id_{V_j}) 
      \eqw l_k \circ \tu_{Vi\times V_j}  $}
\end{prooftree}

\vspace{20pt}

\hrulefill 
\caption{Case $k\neq i,j$ (with $i\ne j$)}
\label{prooftree1}
\end{small}
\end{figure}

\begin{figure}[!h]
\begin{small}

\vspace{20pt}

\flushleft{Proof \pne:} 
\begin{prooftree} 
\AxiomC{\anliuj}
\noLine
\UnaryInfC{$l_i \circ u_j \eqw l_i \circ \tu_{V_j}$}
\LeftLabel{\rnwsubs} 
\UnaryInfC{$l_i \circ u_j \circ \pi_2 \circ (u_i \rtimes \id_{V_j}) 
  \eqw l_i \circ \tu_{V_j} \circ \pi_2 \circ (u_i \rtimes \id_{V_j})$}
    \AxiomC{$\tu_{V_j} \circ \pi_2: V_i\times V_j \to \unit$}
    \RightLabel{\rnwfinal}
    \UnaryInfC{$\tu_{V_j}^\pure \circ \pi_2^\pure \eqw \tu_{1\times V_j}^\pure$}
    \RightLabel{\rnws}
    \UnaryInfC{$\tu_{V_j} \circ \pi_2 \eqs \tu_{1\times V_j}$}
    \RightLabel{\rnsrepl}
    \UnaryInfC{$l_i \circ \tu_{V_j} \circ \pi_2 \eqs l_i \circ \tu_{1\times V_j}$}
    \RightLabel{\rnsw}
    \UnaryInfC{$l_i \circ \tu_{V_j} \circ \pi_2 \eqw l_i \circ \tu_{1\times V_j}$}
  \LeftLabel{\rnwtrans} 
  \BinaryInfC{$l_i \circ u_j \circ \pi_2 \circ (u_i \rtimes \id_{V_j}) 
  \eqw l_i \circ \tu_{1\times V_j} \circ (u_i \rtimes \id_{V_j})$}
\end{prooftree}

\vspace{20pt}

\flushleft{Proof \pnf:} 
\begin{prooftree} 
\AxiomC{$\pi_1: \unit\times V_j \to \unit$}
\LeftLabel{\rnwfinal}
\UnaryInfC{$\tu_{\unit \times V_j}^\pure \eqw \pi_1^\pure$}
\LeftLabel{\rnpa}
\UnaryInfC{$\tu_{\unit \times V_j}^\acc \eqw \pi_1^\acc$}
\LeftLabel{\rnws}
\UnaryInfC{$\tu_{\unit \times V_j} \eqs \pi_1$}
\LeftLabel{\rnssubs}
\UnaryInfC{$\tu_{1\times V_j} \circ (u_i \rtimes \id_{V_j}) 
  \eqs \pi_1 \circ (u_i \rtimes \id_{V_j})$}
    \AxiomC{\anprodmodi}
    \noLine
    \UnaryInfC{$\pi_1 \circ (u_i \rtimes \id_{V_j}) \eqs u_i \circ \pi_1$}
  \LeftLabel{\rnstrans}
  \BinaryInfC{$\tu_{1\times V_j} \circ (u_i \rtimes \id_{V_j}) \eqs u_i \circ \pi_1$}
  \LeftLabel{\rnssubs}
  \UnaryInfC{$l_i \circ \tu_{1\times V_j} \circ (u_i \rtimes \id_{V_j}) 
    \eqs l_i \circ u_i \circ \pi_1$}
  \LeftLabel{\rnsw}
  \UnaryInfC{$l_i \circ \tu_{1\times V_j} \circ (u_i \rtimes \id_{V_j}) 
    \eqw l_i \circ u_i \circ \pi_1$}
\end{prooftree} 

\vspace{20pt}

\flushleft{Proof \png:} 
\begin{prooftree}
\AxiomC{\pne}
    \AxiomC{\pnf}
  \LeftLabel{\rnwtrans}
  \BinaryInfC{$ l_i \circ u_j \circ \pi_2 \circ (u_i \rtimes \id_{V_j}) 
    \eqw l_i \circ u_i \circ \pi_1$}
      \AxiomC{\anliui}
      \noLine
      \UnaryInfC{$l_i \circ u_i \eqw \id_{V_i}$}
      \RightLabel{\rnwsubs}
      \UnaryInfC{$l_i \circ u_i \circ \pi_1 \eqw \pi_1$}
    \LeftLabel{\rnwtrans}
    \BinaryInfC{$ l_i \circ u_j \circ \pi_2 \circ (u_i \rtimes \id_{V_j}) 
      \eqw \pi_1 $}
\end{prooftree} 

\vspace{20pt}

\flushleft{Proof \pnh:} 
\begin{prooftree}
\AxiomC{\anliui}
\noLine
\UnaryInfC{$l_i \circ u_i \eqw \id_{V_i}$}
\LeftLabel{\rnwsubs}
\UnaryInfC{$l_i \circ u_i \circ \pi_1 \circ (\id_{V_i} \ltimes u_j) 
  \eqw \pi_1 \circ (\id_{V_i}\ltimes u_j)$}
    \AxiomC{\anprodpure}
    \noLine
    \UnaryInfC{$\pi_1 \circ (\id_{V_i}\ltimes u_j) \eqw \pi_1$}
    \RightLabel{\rnwsym}
    \UnaryInfC{$\pi_1 \eqw \pi_1 \circ (\id_{V_i}\ltimes u_j)$}
  \LeftLabel{\rnwtrans}
  \BinaryInfC{$l_i \circ u_i \circ \pi_1 \circ (\id_{V_i} \ltimes u_j) 
    \eqw \pi_1$}
\end{prooftree}

\vspace{20pt}

\hrulefill 
\caption{Case $k=i$ (with $i\ne j$)}
\label{prooftree2}
\end{small}
\end{figure}

The diagrams in Figures~\ref{proofdiagram}, 
together with the rules $\rnsw$ and $\rnwtrans$,  
provide a slightly different proof of the weak equations~$(6)_{\st,\ob}$.
In these diagrams we use the derived rule $\rnsfinal$ 
which has been proved in example~\ref{example:toterminal}, 
and (under the same name) 
its consequence $\pi_1 \eqs \tu_{\unit\times X} : \unit\times X \to \unit$. 

\begin{figure}[!h]

\flushleft{Proof \pnd:} 
$$
\xymatrix@R=.5pc{
V_i\times V_j \ar[r]^{u_i \rtimes \id } & 
  \unit \times V_j \ar[r]^{\pi_2} & 
  V_j \ar[r]^{u_j} & 
  \unit \ar[r]^{l_k} & V_k && \\ 
&&  V_j \ar@{=}[u] \ar@{}[urr]|{\eqw} \ar[r]^{\tu} & 
  \unit \ar[r]^{l_k} & V_k \ar@{=}[u] \ar@{}[urr]|{\anliuj} && \\ 
V_i\times V_j \ar@{=}[uu] \ar@{}[urrrr]|{\eqw} \ar[r]^{u_i \rtimes \id } & 
  \unit \times V_j \ar[r]^{\pi_2} & 
  V_j \ar[r]^{\tu} & 
  \unit \ar[r]^{l_k} & V_k \ar@{=}[u] \ar@{}[urr]|{\rnwsubs} && \\ 
& \unit \times V_j \ar@{=}[u] \ar@{}[urr]|{\eqs} \ar[rr]^(.4){\pi_1} && 
  \unit \ar@{=}[u] & \ar@{}[urr]|{\rnsfinal} && \\ 
V_i\times V_j \ar@{=}[uu] \ar@{}[urrrr]|{\eqs} \ar[r]^{u_i \rtimes \id } & 
  \unit \times V_j \ar[rr]^{\pi_1} && 
  \unit \ar[r]^{l_k} & V_k \ar@{=}[uu] \ar@{}[urr]|{\quad\rnssubs,\rnsrepl} && \\ 
V_i\times V_j \ar@{=}[u] \ar@{}[urrr]|{\eqs} \ar[r]^{\pi_1} & 
  V_i \ar[rr]^{u_i} && \unit  \ar@{=}[u] & \ar@{}[urr]|{\anprodmodi} && \\ 
V_i\times V_j \ar@{=}[u] \ar@{}[urrrr]|{\eqs} \ar[r]^{\pi_1} & 
  V_i \ar[rr]^{u_i} && \unit \ar[r]^{l_k} & 
  V_k \ar@{=}[uu] \ar@{}[urr]|{\rnsrepl} && \\ 
& V_i \ar@{=}[u] \ar@{}[urrr]|{\eqw} \ar[rr]^{\tu} && \unit \ar[r]^{l_k} & 
  V_k \ar@{=}[u] \ar@{}[urr]|{\anliuj} && \\ 
V_i\times V_j \ar@{=}[uu] \ar@{}[urrrr]|{\eqw} \ar[r]^{\pi_1} & 
  V_i \ar[rr]^{\tu} && \unit \ar[r]^{l_k} & 
  V_k \ar@{=}[u] \ar@{}[urr]|{\rnwsubs} && \\ 
V_i\times V_j \ar@{=}[u] \ar@{}[urrr]|{\eqs} \ar[rrr]^(.4){\tu} & 
  && \unit \ar@{=}[u] & \ar@{}[urr]|{\rnsfinal} && \\ 
V_i\times V_j \ar@{=}[u] \ar@{}[urrrr]|{\eqs} \ar[rrr]^{\tu} & 
  && \unit \ar[r]^{l_k} & 
  V_k \ar@{=}[uu] \ar@{}[urr]|{\rnsrepl} && \\ 
}$$
\flushleft{Proof \png:} 
$$ \xymatrix@R=.5pc{
V_i\times V_j \ar[r]^{u_i \rtimes \id } & 
  \unit \times V_j \ar[r]^{\pi_2} & 
  V_j \ar[r]^{u_j} & 
  \unit \ar[r]^{l_i} & V_i && \\ 
&&  V_j \ar@{=}[u] \ar@{}[urr]|{\eqw} \ar[r]^{\tu} & 
  \unit \ar[r]^{l_i} & V_i \ar@{=}[u] \ar@{}[urr]|{\anliuj} && \\ 
V_i\times V_j \ar@{=}[uu] \ar@{}[urrrr]|{\eqw} \ar[r]^{u_i \rtimes \id } & 
  \unit \times V_j \ar[r]^{\pi_2} & 
  V_j \ar[r]^{\tu} & 
  \unit \ar[r]^{l_i} & V_i \ar@{=}[u] \ar@{}[urr]|{\rnwsubs} && \\ 
& \unit \times V_j \ar@{=}[u] \ar@{}[urr]|{\eqs} \ar[rr]^(.4){\pi_1} && 
  \unit \ar@{=}[u] & \ar@{}[urr]|{\rnsfinal} && \\ 
V_i\times V_j \ar@{=}[uu] \ar@{}[urrrr]|{\eqs} \ar[r]^{u_i \rtimes \id } & 
  \unit \times V_j \ar[rr]^{\pi_1} && 
  \unit \ar[r]^{l_i} & V_i \ar@{=}[uu] \ar@{}[urr]|{\quad\rnssubs,\rnsrepl} && \\ 
V_i\times V_j \ar@{=}[u] \ar@{}[urrr]|{\eqs} \ar[r]^{\pi_1} & 
  V_i \ar[rr]^{u_i} && \unit  \ar@{=}[u] & \ar@{}[urr]|{\anprodmodi} && \\ 
V_i\times V_j \ar@{=}[u] \ar@{}[urrrr]|{\eqs} \ar[r]^{\pi_1} & 
  V_i \ar[rr]^{u_i} && \unit \ar[r]^{l_i} & 
  V_i \ar@{=}[uu] \ar@{}[urr]|{\rnsrepl} && \\ 
& V_i \ar@{=}[u] \ar@{}[urrr]|{\eqw} \ar[rrr]^(.4){\id} &&& 
  V_i \ar@{=}[u] \ar@{}[urr]|{\anliui} && \\ 
V_i\times V_j \ar@{=}[uu] \ar@{}[urrrr]|{\eqw} \ar[rrrr]^(.4){\pi_1} &&&& 
  V_i \ar@{=}[u] \ar@{}[urr]|{\rnwsubs} && \\
}$$ 
\flushleft{Proof \pnh:} 
$$
\xymatrix@R=.5pc{
V_i\times V_j \ar[r]^{\id \ltimes u_j} & 
  V_i\times \unit \ar[r]^{\pi_1} & 
  V_i \ar[r]^{u_i} & 
  \unit \ar[r]^{l_i} & V_i && \\ 
&& V_i \ar@{=}[u] \ar@{}[urr]|{\eqw} \ar[rr]^(.4){\id} && 
  V_i \ar@{=}[u] \ar@{}[urr]|{\anliui} && \\ 
V_i\times V_j \ar@{=}[uu] \ar@{}[urrrr]|{\eqw} \ar[r]^{\id \ltimes u_j} & 
  V_i\times \unit \ar[rrr]^{\pi_1} &&& 
  V_i \ar@{=}[u] \ar@{}[urr]|{\rnwsubs} && \\ 
V_i\times V_j \ar@{=}[u] \ar@{}[urrrr]|{\eqw} \ar[rrrr]^(.4){\pi_1} &&&& 
  V_i \ar@{=}[u] \ar@{}[urr]|{\anprodpure} && \\  
}$$ 

\vspace{10pt}

\hrulefill 
\caption{Proofs as diagrams}
\label{proofdiagram}
\end{figure}

\medskip %

Equation $(3)$ is expressed in the decorated logic as:
\[ (3)_\st \; \forall i\in\Loc,\; 
u_i^\modi\circ \pi_2^\pure \circ (u_i\rtimes \id_{V_i})^\modi \eqs
u_i^\modi \circ \pi_2^\pure \]
which is equivalent to:
\[ (3)_{\st,\ob} \;
\forall k\in\Loc, \forall i\in\Loc,\; 
l_k^\acc \circ u_i^\modi \circ \pi_2^\pure \circ (u_i\rtimes \id_{V_i})^\modi \eqw
l_k^\acc \circ u_i^\modi \circ \pi_2^\pure \] 
We can again split the proof in two cases, with proof trees 
similar to those for equations~$(6)_{\st,\ob}$:
\begin{enumerate} 
\item
  When $k\neq i$, both sides reduce (in the $\eqw$ sense) to $l_k\circ
  \tu_{V_i \times V_j}$. 
\item
  When $k = i$, both sides reduce (in the $\eqw$ sense) to $\pi_2$. 
\end{enumerate}

\newpage\null
\newpage\null
\newpage

\subsection{Exceptions}
\label{app:exceptions}

Dually, we get the decorated equations for exceptions. 
In the decorated logic for exceptions, 
the \emph{coproduct} of two objects $A$ and $B$ 
is an object $A+B$ with two pure morphisms: the coprojections
$\copi_1^\pure:A \to A+B$ and $\copi_2^\pure:B \to A+B$, 
which satisfy the usual categorical coproduct property 
with respect to the pure morphisms. 
So, as usual, the coprojections 
$\copi_1^\pure:A \to A+\empt$ and $\copi_2^\pure:B \to \empt+B$
are isomorphisms. 
The \emph{semi-pure coproduct} of a pure morphism $f^\pure:A\to C$ 
and a catcher $g^\ctc:B\to D$ is a catcher 
$(f\lplus g)^{\ctc}:A+B\to C+D$ 
(for simplicity we still use the symbol $+$) 
which is characterized, up to $\eqs$, by 
the following decorated version of the coproduct property: 
\begin{align*}
(f \lplus g)^\ctc \circ \copi_1^\pure & \eqw \copi_1^\pure \circ f^\pure \\
(f \lplus g)^\ctc \circ \copi_2^\pure & \eqs \copi_2^\pure \circ g^\ctc \\ 
\end{align*}
$$ 
\xymatrix@C=4pc{
  A \ar[r]^{f^\pure} \ar[d]_{\copi_1^\pure} & C \ar[d]^{\copi_1^\pure} \\
  A + B \ar[r]^{(f \lplus g)^\ctc} 
  \ar@{}[ur]|{\eqw}
   & C + D \\
  B \ar[r]_{g^\ctc} \ar[u]^{\copi_2^\pure} \ar@{}[ur]|{\eqs} & 
   D \ar[u]_{\copi_2^\pure} \\ 
}
$$ 

Now, the equations $(6)_\exc$ and $(3)_\exc$, 
dual to equations $(6)_\st$ and $(3)_\st$, 
can be proved in the dual way. 
$$ \begin{array}{ll}
(6)_\exc  
& \forall i\neq j \in \Const ,\; 
  (c_i\rplus \id_{P_j})^\ctc \circ \copi_2^\pure \circ c_j^\ctc \eqs 
    (\id_{P_i} \lplus c_j)^\ctc \circ \copi_1^\pure \circ c_i^\ctc \\ 
(3)_\exc  
& \forall i\in\Const ,\; 
(c_i\rplus \id_{P_i})^\ctc \circ \copi_2^\pure \circ c_i^\ctc \eqs
\copi_2^\pure \circ c_i^\ctc \\ 
\end{array} $$

Proposition~\ref{proposition:exceptions-two} can be proved 
by encapsulating these equations, 
if it is assumed that the coproducts in the decorated logic
are coproducts for the propagators.
This means that for each propagators $g^\acc:A\to C$ and $h^\acc:B\to C$ 
there is a propagator $\cotuple{g\;|\;h}^\acc:A+B \to C$
which is characterized, up to $\eqs$, by: 
\begin{align*}
\cotuple{g\;|\;h}^\acc \circ \copi_1^\pure & \eqs g^\acc \\
\cotuple{g\;|\;h}^\acc \circ \copi_2^\pure & \eqs h^\acc \\ 
\end{align*}
Proposition~\ref{proposition:exceptions-two} states that:
$$ \begin{array}{ll}
(6)_{\exc,\encaps}  
& \forall i\neq j \in \Const , 
  \try{f}{\catchrec{i}{g}{\catch{j}{h}}} \eqs 
  \try{f}{\catchrec{j}{h}{\catch{i}{g}}} \\
(3)_{\exc,\encaps}  
& \forall i\in\Const ,\; 
  \try{f}{\catchrec{i}{g}{\catch{i}{h}}} \eqs 
  \try{f}{\catch{i}{g}} \\ 
\end{array} $$ 
where, according to definition~\ref{definition:exceptions-handle}: 
$$ \begin{array}{l}
\forall i, j \in \Const ,\; 
  \try{f}{\catchrec{i}{g}{\catch{j}{h}}} = 
  \toppg{( 
    \cotuple{\id \;|\; 
      \cotuple{g \;|\; 
        h \circ c_j} 
      \circ c_i} 
    \circ f )} \\ 
\forall i\in\Const ,\; 
  \try{f}{\catch{i}{g}} = 
  \toppg{( 
    \cotuple{\id \;|\; 
      g \circ c_i} 
    \circ f )} \\ 
\end{array} $$ 
  $$ 
  \xymatrix@C=3pc@R=1.5pc{
  X \ar[r]_(.6){f^\ppg} 
    \ar@<1ex>@/^4ex/[rrrr]^{(\try{f}{\catchrec{i}{g}{\catch{j}{h}}})^\ppg}_{\eqw} & 
    Y \ar[rrr]_(.4){\;[\id|[g|h \circ c_j]\circ c_i]^\ctc\;} &&& Y \\ 
  & \empt \ar[u]^{\cotu^\pure} \ar[r]^{c_i^\ctc} & 
    P_i  \ar@{}[u]|{\eqs} \ar@/_/[rru]_(.4){\;[g|h \circ c_j]^\ctc\;} 
      && \\
  && \empt \ar[u]^{\cotu^\pure} \ar[r]^{c_j^\ctc} & 
    P_j  \ar@{}[u]|{\eqs} \ar@/_/[ruu]_(.4){\;h^\ppg\;} \\ 
  } \quad 
  \xymatrix@C=3pc@R=1.5pc{ 
  X \ar[r]_(.6){f^\ppg} \ar@<1ex>@/^3ex/[rrr]^{(\try{f}{\catch{i}{g}})^\ppg}_{\eqw} & 
    Y \ar[rr]_{[\id\,|\,g\circ c_i]^\ctc} && Y \\ 
  & \empt \ar[u]_{\cotu^\pure} \ar[r]_{c_i^\ctc} & 
    P_i \ar@/_/[ru]_{g^\ppg} \ar@{}[u]|{\eqs} & \\
  } $$

\textit{Proof of proposition~\ref{proposition:exceptions-two}.}
It is easy to check that 
 $$ \cotuple{g \;|\; h \circ c_j}
  \eqs \cotuple{g \;|\; h} \circ (\id_{P_i} \lplus c_j) $$
then it follows from $(6)_\exc$ and $(3)_\exc$ that 
$$ \begin{array}{l}
\forall i\neq j  ,\; 
\cotuple{g \;|\; h \circ c_j} \circ c_i \eqs 
\cotuple{h \;|\; g \circ c_i} \circ c_j \\
\forall i ,\; 
\cotuple{g \;|\; h \circ c_i} \circ c_i \eqs 
g \circ c_i \\ 
\end{array} $$ 
which implies $(6)_{\exc,\encaps}$ and $(3)_{\exc,\encaps}$,
as required.
\qed

\renewcommand{\arraystretch}{1}

\end{document}